\documentclass[sigconf]{acmart}
\AtBeginDocument{%
  \providecommand\BibTeX{{%
    \normalfont B\kern-0.5em{\scshape i\kern-0.25em b}\kern-0.8em\TeX}}}

\copyrightyear{2025}
\acmYear{2025}
\setcopyright{cc}
\setcctype{by}
\acmConference[FAccT '25]{The 2025 ACM Conference on Fairness, Accountability, and Transparency}{June 23--26, 2025}{Athens, Greece}
\acmBooktitle{The 2025 ACM Conference on Fairness, Accountability, and Transparency (FAccT '25), June 23--26, 2025, Athens, Greece}\acmDOI{10.1145/3715275.3732050}
\acmISBN{979-8-4007-1482-5/2025/06}

\usepackage{cleveref}
\usepackage[shortlabels]{enumitem}
\usepackage{multirow}
\usepackage{siunitx}
\usepackage{stfloats}
\usepackage{subcaption}

{
\setlength{\topsep}{0.5em}
\newtheoremstyle{theorembf}
  {\topsep}
  {\topsep}
  {\itshape}
  {0pt}
  {\bfseries\scshape}
  {.}
  { }
  {\thmname{#1}\thmnumber{ #2}\textnormal{\thmnote{ (#3)}}}
\theoremstyle{theorembf}
\newtheorem{rsq}{Research Question }%

\newtheoremstyle{theoremit}
  {\topsep}
  {\topsep}
  {\itshape}
  {0pt}
  {\scshape}
  {.}
  { }
  {\thmname{#1}\thmnumber{ #2}\textnormal{\thmnote{ (#3)}}}
\theoremstyle{theoremit}

}
\crefname{rsq}{research question}{research questions}
\Crefname{rsq}{Research Question}{Research Questions}
\crefname{hyp}{hypothesis}{hypotheses}
\Crefname{hyp}{Hypothesis}{Hypotheses}

\begin{document}

\title[Missing Pieces]{Missing Pieces: How Do Designs that Expose Uncertainty Longitudinally Impact Trust in AI Decision Aids? An In Situ Study of Gig Drivers}

\author{Rex Chen}
\email{rexc@cmu.edu}
\orcid{0000-0002-1620-0440}
\affiliation{%
  \institution{Software \& Societal Systems Department, School of Computer Science, Carnegie Mellon University}
  \streetaddress{4665 Forbes Ave}
  \city{Pittsburgh}
  \state{Pennsylvania}
  \country{USA}
  \postcode{15213}
}
\author{Ruiyi Wang}
\authornote{Work completed at Carnegie Mellon University}
\email{ruiyiwang@ucsd.edu}
\affiliation{%
  \institution{Department of Computer Science and Engineering, University of California, San Diego}
  \streetaddress{3235 Voigt Dr}
  \city{La Jolla}
  \state{California}
  \country{USA}
  \postcode{92093}
}
\author{Norman Sadeh}
\email{sadeh@cs.cmu.edu}
\orcid{0000-0003-4829-5533}
\affiliation{%
  \institution{Software \& Societal Systems Department, School of Computer Science, Carnegie Mellon University}
  \streetaddress{4665 Forbes Ave}
  \city{Pittsburgh}
  \state{Pennsylvania}
  \country{USA}
  \postcode{15213}
}
\author{Fei Fang}
\email{feifang@cmu.edu}
\orcid{0000-0003-2256-8329}
\affiliation{%
  \institution{Software \& Societal Systems Department, School of Computer Science, Carnegie Mellon University}
  \streetaddress{4665 Forbes Ave}
  \city{Pittsburgh}
  \state{Pennsylvania}
  \country{USA}
  \postcode{15213}
}

\renewcommand{\shortauthors}{Chen et al.}

\begin{abstract}
    Decision aids based on artificial intelligence (AI) induce a wide range of outcomes when they are deployed in uncertain environments. In this paper, we investigate how users' trust in recommendations from an AI decision aid is impacted over time by designs that expose uncertainty in predicted outcomes. Unlike previous work, we focus on gig driving --- a real-world, repeated decision-making context. We report on a longitudinal mixed-methods study ($n=51$) where we measured gig drivers' trust as they interacted with an AI-based schedule recommendation tool. Our results show that participants' trust in the tool was shaped by both their first impressions of its accuracy and their longitudinal interactions with it, and that task-aligned framings of uncertainty improved trust by allowing participants to incorporate uncertainty into their decision-making processes. Additionally, we observed that trust depended on their characteristics as drivers, underscoring the need for more in situ studies of AI decision aids.
\end{abstract}

\ccsdesc[500]{Human-centered computing~User studies}
\ccsdesc[500]{Human-centered computing~Empirical studies in HCI}
\ccsdesc[500]{Human-centered computing~Empirical studies in collaborative and social computing}
\ccsdesc[300]{Applied computing~Transportation}
\ccsdesc[300]{Human-centered computing~HCI theory, concepts and models}
\keywords{Human-AI trust, human-AI compliance, human-AI interaction, AI design, AI decision aids, gig driving, recommendation systems, longitudinal studies}


\maketitle

\section{Introduction}
\label{sec:intro}
The fairness, accountability, and transparency of algorithms are inextricably intertwined with the extents to which people are willing to interact with them. Accordingly, the FAccT community has increasingly focused on studying trust in algorithmic systems, especially those based on \emph{artificial intelligence} (AI) \cite{Ferrario2022,Jacovi2021,Kim2023,Knowles2021,Liao2022b}. In this paper, we study how people \emph{trust} and \emph{rely} on \emph{AI decision aids}. AI decision aids function by (1) recommending decisions and (2) predicting how good the outcomes of following those decisions will be. Over recent years, AI decision aids have been adopted at an increasing rate in a number of application domains \cite{Milana2021,McConvey2023,Riahi2021,Skjuve2023,Wang2021a,Zavrsnik2019}. However, because AI decision aids are used in the context of people and other sociotechnical systems, considerable \emph{uncertainty} in decision outcomes can exist \cite{Russell2017,Macrae2021}. When this uncertainty impacts the predictability of AI decision aids, users' trust in the decision aid can be eroded \cite{Jacovi2021,Solberg2022}.

Prior literature on trust in AI decision aids under uncertainty can be organised into two complementary lines of work. One line of work has studied specific factors that influence trust in AI decision aids, using laboratory experiments in simulated, single-shot, and low-stakes scenarios that require limited domain expertise \cite{Bansal2021,Cai2019,Jacobs2021,Kocielnik2019,Zhang2020b}. However, context is an important factor for trust in AI \cite{Kim2023,Solberg2022}. Therefore, the contrived nature of these experiments limits their generalisability to the real-world use contexts of AI decision aids. Another line of work has studied trust in real-world AI decision aids \cite{Beede2020,Kim2023,Wang2021a}, using observational, qualitative studies to assess how existing users interact with these decision aids. These studies are not quantitative assessments of design factors and provide limited insight into new, trustworthy AI decision aids can be designed.

In this paper, we provide a deeper exploration of trust in AI decision aids by combining the strengths of these two lines of work. We contribute the first in situ study of how exposing the uncertainty of an AI decision aid has longitudinal impacts on users' trust and reliance on the decision aid. Using the domain of \emph{gig driving} --- in which drivers use their personal vehicles to fulfil ridesourcing and food delivery requests from platforms such as Uber, Lyft, DoorDash, and Instacart --- as a testbed, we study trust in a real-world, medium-to-high-stakes decision-making scenario where users have existing expertise. Specifically, we comparatively evaluate different designs that expose the potential for misprediction in an AI decision aid. We address the following research questions:

\begin{rsq}
    \label{rsq:accuracy}
    How do users' trust and reliance on an AI decision aid depend longitudinally on their perception of its predictive accuracy?
\end{rsq}
\begin{rsq}
    \label{rsq:uncertainty}
    How do different designs that expose the inherent uncertainty in predictive performance impact users' trust and reliance on an AI decision aid?
\end{rsq}

We addressed these questions by conducting a longitudinal user study where $n = 51$ gig drivers used an AI-based schedule recommendation tool. By measuring the trust and reliance of participants over repeated interactions, we tested the effects of exposing uncertainty in the tool's predictions through range-based earnings estimates and hedging text. Our quantitative and qualitative findings show that participants' initial perceptions of the tool's accuracy improved their trust in it over time. In addition, range-based uncertainty not only improved trust and reliance in single-shot settings, but also strengthened it over repeated interactions; meanwhile, hedging text had the opposite effect.

\section{Related Work}
\label{sec:related}

\subsection{Trust in AI}
\label{sec:related:trust}
A lack of uniformity exists in the human-AI interaction literature on how to define and evaluate trust in AI systems \cite{Ueno2022}. We follow \citet{Mayer1995} in operationalising the constructs of \emph{trust} and \emph{reliance}: in the context of AI decision aids, they hypothesise that trust is ``the willingness of a person to be vulnerable to the actions of an [AI decision aid], based on the expectation that it will perform a [decision-making task] important to the trustor'', and that reliance is the external, behavioural expression of that internal attitude \cite{Solberg2022}. One point of broad consensus in the literature is that the design of AI systems impacts expectations of their performance in decision-making, and thus the trust of their users \cite{Burgess2023,Danner2020,Jacobs2021,Kocielnik2019,Kunkel2019,Scharowski2023,Yu2019}.

However, the majority of this work has been based on controlled laboratory experiments. Compared to real-world use contexts of AI decision aids, such experiments have two main limitations: (1)~They are \emph{single-shot}, involving only a single session of AI use with no temporal separation between decision points \cite{Ueno2022}. Several studies have compared \emph{two} trust measurements \cite{Mou2017a,Mou2017b,Kunkel2019}, but these studies still provide a limited perspective on the evolution of trust dynamics. (2)~They are \emph{low-stakes}, involving contrived or hypothetical decision-making scenarios in which an element of \emph{risk} and thus vulnerability is largely absent \cite{Kim2023}. Some work has studied AI decision aids for decision-making domains that entail high stakes in the real world, such as the medical \cite{Burgess2023,Cai2019,Beede2020,Jacobs2021,Wang2021a,Sivaraman2023,Wysocki2023,Yang2023} and financial \cite{Prabhudesai2023,Schoeffer2022,Scharowski2023} contexts. However, for practical and ethical reasons, participants in these studies cannot receive feedback from the real world for their decisions. A minority of work has evaluated trust in AI decision aids within their real-world contexts \cite{Beede2020,Kim2023,Wang2021a,Widder2021}, but these have been limited to observational studies that did not compare multiple designs.

\emph{Uncertainty} encapsulates sources of variability that make it difficult for users to reason about the outcomes of relying on an AI, thus increasing the risk of this reliance \cite{Stuck2021}. Various experimental studies have tested the effects of designs that make users more aware of the presence and impact of uncertainty on AI \cite{Bansal2021,Cau2023,Kocielnik2019,Naiseh2023,Wang2021b,Yang2024}. This work has parallels to \emph{trust calibration} in the AI explainability literature: giving users realistic expectations regarding when and why AI systems may or may not perform well \cite{Bhatt2021,Tomsett2020,Zhang2020b}. In addition to the lack of real-world context for these experiments, one thus far underexplored design is the use of \emph{lexical hedging}: verbiage that expresses uncertainty. \citet{Kim2024} assessed the effects of lexical hedging in a large language model's medical answers on trust but not on reliance. \citet{Zhang2022a} measured trust and reliance on an AI decision aid with lexical hedging for a contrived shape identification task. We perform a real-world evaluation of two designs for presenting uncertainty in an AI decision aid: presenting scalar ranges for estimates, and qualifying estimates with lexical hedging.

\subsection{Gig Work and Gig Driving}
\label{sec:related:gigs}
Gig work offers workers the flexibility and autonomy to choose when and where they want to work \cite{Lee2015}. However, this autonomy is hampered by the opacity of platforms' assignment, pricing, and evaluation mechanisms \cite{Griesbach2019,Lee2015,Singh2023,Tan2021,Wood2019}. In particular, gig driving platforms impose information asymmetry \cite{Maffie2023,Rosenblat2016} by dynamically varying their compensation mechanisms \cite{Lee2015,Zhang2022b}. For drivers, this lack of visibility leads to difficulties in planning \cite{Griesbach2019} and inequities in revenue \cite{Miao2023}. This makes gig driving an exemplary context to assess the impact of uncertainty, repeated interactions, and moderate financial stakes on AI decision aids.

Within the gig driving setting, the most relevant prior work to our paper consists of studies that have designed predictive and prescriptive tools for gig drivers \cite{Battifarano2019,Khan2022,Zhang2022b,Zhang2023,Zhang2024}. Among these, \citet{Battifarano2019} and \citet{Khan2022} focused on evaluating the predictive accuracy of their systems and did not include a user design component; \citet{Zhang2022b,Zhang2023} conducted formative studies that designed tools in collaboration with drivers but did not deploy them to evaluate users' trust and reliance; and \citet{Zhang2024} tested the effects of different AI decision aid designs that displayed uncertainty, albeit using simulated taxi trip data and focusing on the decisions --- not attitudes --- of participants.

\section{AI-Based Schedule Recommendation for Gig Driving}
\label{sec:tool}
In this section, we describe an AI-based schedule recommendation tool for gig drivers. Like other AI decision aids \cite{Solberg2022}, this tool (1) recommends a set of decisions (i.e. a schedule) to achieve a set objective (see \Cref{sec:tool:decisionaid}), and (2) predicts the outcomes (i.e. estimated earnings) of following those decisions. We use this tool as an exemplary AI decision aid to study the longitudinal relationship between the framing of uncertainty in outcomes and trust. The design of this tool was conceptualised and refined through a formative pilot study, which we describe in \Cref{sec:app:pilot}.

\subsection{Decision Aid Design}
\label{sec:tool:decisionaid}
We focus on one aspect of planning for gig drivers: choosing \emph{when} to work, subject to their constraints and preferences. There is considerable variation in drivers' habits along this dimension \cite{Berger2020,Lee2015,Ma2019}. Choosing \emph{where} to work is another key aspect of planning \cite{Zhang2023}, but we limited our study of uncertainty in this multi-objective problem to a single dimension. 

Our tool consists of two modules. First, an \textbf{estimation module} prospectively predicts $e_{ij}$, the earnings that drivers can expect during a specific hour $j$ on a specific weekday $i$. These could be computed by a machine learning model or averaged from historical data. Second, for each driver, a \textbf{scheduling module} uses the estimated earnings and the driver's constraints as inputs to produce an optimal set of working times. To do so, it solves a constrained optimisation problem to set \emph{variables} $x_{ij}$ to 1 or 0, denoting whether the driver is recommended to work in time slot $(i,j)$.

\begin{itemize}
    \item Some drivers wish to maximise their earnings while minimising their driving hours. For these drivers, we maximise the \emph{objective function}: the sum of the estimated earnings $e_{ij}$ for all recommended time slots $(i,j)$ from the estimation module, i.e. $\sum_{i,j} e_{ij} x_{ij}$. 
    
    To set the \emph{constraints}, we disallow time slots when the driver is not available, i.e. $x_{ij} \leq a_{ij}$ where $a_{ij}$ is an indicator of whether the driver is available during time slot $(i,j)$. We also place an upper bound on the total hours of recommended time slots per day by $b_i$, and per week by $b_{tot}$, i.e. $\sum_{j} x_{ij} \leq b_i, \forall i; \sum_{i,j} x_{ij} \leq b_{tot}$. 

    \item Some drivers who value earnings to a greater extent set minimum targets for their hourly or daily earnings instead of restricting their driving hours. For these drivers, we minimise the \emph{objective function}: the total hours of recommended time slots throughout the week, i.e. $\sum_{i,j} x_{ij}$.
    
    To set the \emph{constraints}, we disallow time slots when the driver is not available, i.e. $x_{ij} \leq a_{ij}$. We also place a lower bound on the estimated earnings per day by $c_i$, and per week by $c_{tot}$, i.e. $\sum_{j} e_{ij} x_{ij} \geq c_i, \forall i; \sum_{i,j} e_{ij} x_{ij} \geq c_{tot}$. 
\end{itemize}

\subsection{Interface Design}
\label{sec:tool:design}
Next, we also designed a front-end interface that would allow drivers to interact with the tool. The interface was implemented as a HTML/CSS/JavaScript website using Django 4.1 \cite{Django2023} and a PostgreSQL database. To mitigate potential biases, we designed our tool to be visually generic and distinct from apps or websites associated with any gig platforms.

First, a \textbf{constraint page} (\Cref{fig:tool:constraint} in \Cref{sec:app:tool}) elicits the constraints needed for the scheduling module. It prompts users to select an optimisation objective: whether to maximise earnings or minimise hours on a daily or weekly basis ($b_i$, $b_{tot}$, $c_i$, and $c_{tot}$). To maximise perceived control over the tool, we allowed users to choose these options freely rather than assigning them as conditions. The page also elicits hourly availability information ($a_{ij}$). Unlike \citet{Zhang2023}, we account for the fact that users' availability may change between days.

Second, a \textbf{schedule page} (\Cref{sec:tool:conditions}) presents a tabular schedule to the user. The optimal schedule is shown by highlighting the recommended time slots, i.e. the ones that lead to the highest earnings. Again, in the interests of maximising perceived control over the tool, we allowed users to revisit the constraint page until they were satisfied with the schedule.

\subsection{Interface Conditions}
\label{sec:tool:conditions}
\begin{figure*}[ht]
    \centering
    \includegraphics[height=0.9\textheight]{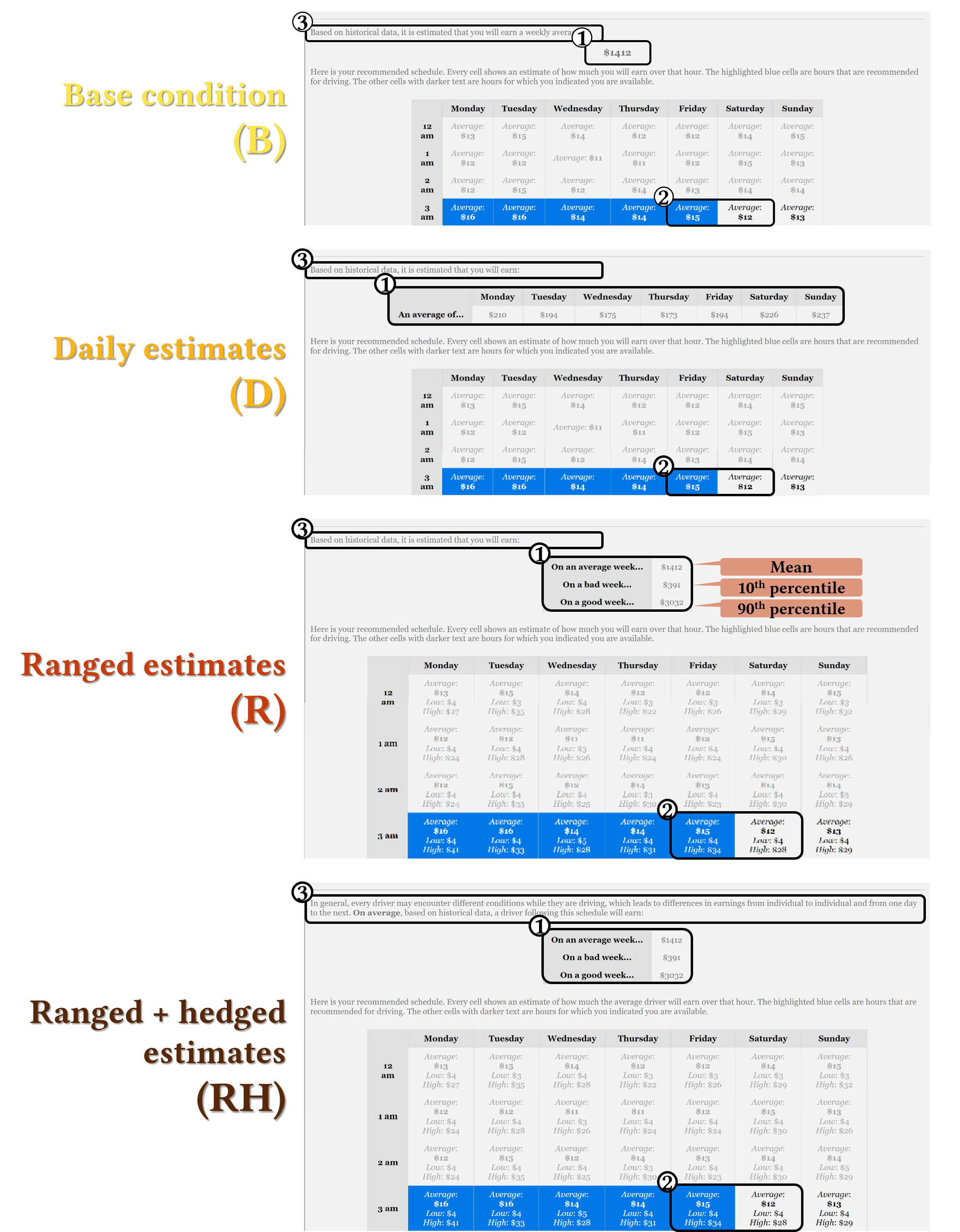}
    \caption{Comparison of the four design conditions for the earnings estimates and recommended schedules on the schedule page, with abbreviations following \Cref{sec:tool:conditions}. Boxes highlight differences between conditions in three areas: (1) weekly earnings estimates, (2) hourly earnings estimates, and (3) textual description of estimates. Full screenshots are shown in \Cref{sec:app:tool}.}
    \label{fig:conditions}
    \Description{Screenshots of a box titled "Shift Recommendation Tool" on the schedule page. Condition (B) is shown at the top. An estimated weekly earning of \$1412 is highlighted with box 1. A recommended schedule shows average earnings for each weekday and each hour. Two cells reading "Average: \$15" and "Average: \$12" are highlighted with box 2. Before the weekly estimate of \$1412, box 3 highlights text reading "Based on historical data, it is estimated that you will earn a weekly average of". The subsequent three subfigures show variants of this page. For box 1 (weekly earnings), Condition (D) displays a table showing "An average of..." with \$210 for Monday, \$194 for Tuesday, etc.; Conditions (R) and (RH), a table showing "On an average week: \$1412", "On a bad week: \$391", and "On a good week: \$3032". For box 2 (hourly earnings), Conditions (R) and (RH) display cells reading "Average: \$15, Low: \$4, High: \$34" and "Average: \$12, Low: \$4, High: \$28". For box 3 (textual description), Condition (RH) displays "In general, every driver may encounter different conditions while they are driving, which leads to differences in earnings from individual to individual and from one day to the next. On average, based on historical data, a driver following this schedule will earn".}
    \vspace{-1em}
\end{figure*}

The schedule page uses the outputs of the estimation module to predict how much a driver following the recommended schedule would make per hour and per week. However, uncertainty inherently exists in these predictions, as they are based on historical data, and their realisation is contingent upon which gigs are offered to and accepted by drivers. To address \Cref{rsq:uncertainty}, we varied the design of the schedule page between four conditions (\Cref{fig:conditions}):

\begin{enumerate}[(1)]
    \item[(B)] \textbf{Base condition}. Users were only shown their mean estimated earnings for the week and for each hour in the week.
    
    \item[(D)] \textbf{Daily estimates}. To assess the effect of introducing additional information irrelevant to uncertainty, users were shown their mean estimated earning for each day instead of their mean estimated weekly earning. As in (B), the schedule still showed mean estimated earnings for each hour.
    
    \item[(R)] \textbf{Ranged estimates}. To assess the effect of exposing uncertainty through range-based estimates (similar to \citet{Prabhudesai2023}), users were shown mean, pessimistic, and optimistic estimates for hourly and weekly earnings. 
    
    \item[(RH)] \textbf{Ranged and hedged estimates}. To assess the effect of exposing uncertainty through lexical hedging (similar to \citet{Kim2024} and \citet{Zhang2022a}), the textual description of the estimates was changed from (R). Instead of ``Based on historical data, it is estimated that you will earn'', (RH) states ``On average, based on historical data, a driver following this schedule will earn''.
\end{enumerate}

\section{Longitudinal User Study Design}
\label{sec:userstudy}
\begin{figure*}[ht]
    \centering
    \includegraphics[trim={0 7cm 0 0},clip,width=0.9\linewidth]{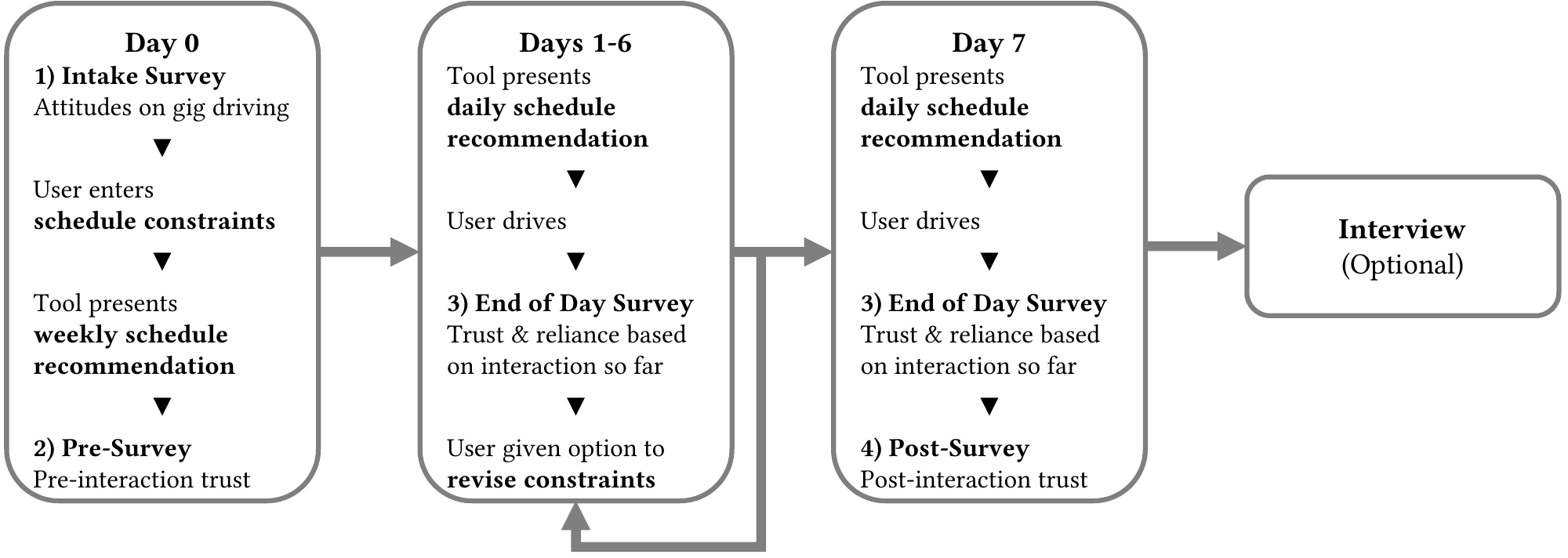}
    \caption{Flow of activities for the longitudinal user study. To be compensated, participants needed to complete Day 0 activities.}
    \label{fig:flow}
    \Description{Flow chart showing bulleted lists of study activities. On Day 0, the Intake Survey is followed by use of the tool to generate the weekly schedule, along with the Pre-Survey as a first measurement of trust. On Days 1-6, the participant receives a daily schedule, drives as usual, and completes the End-of-Day Survey. This repeats until Day 7, which is similar except the End-of-Day Survey is followed by the Post-Survey. This is followed by the optional exit interview.}
\end{figure*}

To address \Cref{rsq:accuracy}, we conducted a longitudinal, in situ user study in which gig drivers repeatedly interacted with our schedule recommendation tool, and we measured their trust and reliance over these interactions. Our participants used the tool for 7 days over a 14-day period, with the longer time window meant to accommodate variability in participants' availability. The methodology for this study was approved by our Institutional Review Board (IRB). \Cref{fig:flow} illustrates the flow of the user study; we detail each day's study activities in \Cref{sec:userstudy:activities}. 

\vfill

Based on a pilot conducted with 7 participants in August 2023, we determined that the Intake Survey, Pre-Survey, and tool interaction on Day 0 took an average of 14 minutes and 27 seconds, the End-of-Day Survey took an average of 2 minutes and 23 seconds per day, and the Post-Survey took an average of 2 minutes and 21 seconds. Based on \citet{Berkel2021}'s recommendation of micro-compensation, this led us to set the compensation as an Amazon gift card with \$6 for the Day 0 surveys, \$2 for each daily survey, and a \$20 completion bonus (\$40 for full study completion). We made one payment upon study completion or the passage of 14 days.

\subsection{Participants and Data Sources}
\label{sec:userstudy:participants}
Participants were recruited from the user base of Gridwise, a mobile assistant app for gig drivers, in September 2023. We chose to recruit from this user base to access a relatively large and diverse sample of both historical data and participants. Gridwise distributed recruitment messages to users who (1) had completed at least one gig in DoorDash, Grubhub, Instacart, Lyft, Uber, or Uber Eats over the month preceding recruitment, and (2) resided in one of the four cities with the historical data used to generate the tool's earnings estimates: Los Angeles, New York, Chicago, and Houston. These were the platforms and cities for which historical data was available.

Accordingly, we generated estimates using gig data from August 2023 in each of these four cities. For each city, our data included approximately 100\,000--300\,000 gig records distributed evenly across times and weekdays. Hourly earnings were estimated by the mean of what drivers historically earned in this slot, filtered to the participant's city and platforms. We used this static estimator to focus on the effects of exposing uncertainty in the estimates. Accordingly, we did not emphasise to participants that the schedule page was generated using AI or optimisation, and did not provide any additional information about the data used to generate the estimates.

\subsection{User Study Activities}
\label{sec:userstudy:activities}
\paragraph{Day~0: Pre-Interaction} 
Participants received a link to the study website from a recruitment message distributed by email. After the consent form, they completed the first of four surveys, the \textbf{Intake Survey} (\Cref{sec:app:survey:1}). This 12-question survey asked about their needs and motivations as gig drivers, along with demographics. 

Next, participants were directed to interact with the tool, which we displayed in an iframe to mitigate response bias \cite{Dell2012,Ko2015}. They entered their constraints on the constraint page, and received the tool's recommended schedule for the entire week on the schedule page. Participants were assigned to one of the four conditions for the schedule page (\Cref{sec:tool:conditions}) uniformly at random, such that each condition had an approximately equal number of participants. 

Lastly, participants completed the second of four surveys, the \textbf{Pre-Survey} (\Cref{sec:app:survey:2}), which was a 5-question survey measuring trust before interaction with the tool (\Cref{sec:quant:instrument}).

\paragraph{Days~1--7: Interaction} 
Next, participants began their 7 days of interaction with the tool, beginning on the next day of the week for which they indicated they were available to drive. 

On each day, participants first received their recommended schedule for that day, sent via an email scheduled for 30 minutes before the start of their indicated availability. Thus, the tool's outputs were displayed right as they were deciding their driving schedules. During the day, participants independently made decisions about their driving activity; we emphasised that compliance with the recommended schedule was not a condition of full participation.

At the end of each participant's indicated for the day availability, a second scheduled email sent them a link to the \textbf{End-of-Day Survey} (\Cref{sec:app:survey:3}). This was an 8-question survey that measured their trust in the tool for that day, and their intention to rely on the tool for the next day (\Cref{sec:quant:instrument}). If the participant intended to continue relying on the tool, they were then presented with a daily variant of the schedule page. Here, they could review the recommended schedule for the following day, and revise their constraints for the day as desired. Updated schedules were generated by fixing the recommended time slots for previous days using equality constraints and then re-solving the optimisation problem. However, if the participant intended to pause their interaction for one day, an email was sent on the next day, which prompted them to either review the next day's schedule or to pause for an additional day.

\paragraph{Day~7: Post-Interaction} 
On the final day, we removed the last question measuring reliance from the End-of-Day Survey, and added the \textbf{Post-Survey} (\Cref{sec:app:survey:4}). This was a 10-question survey that retrospectively measured participants' trust and distrust in the tool over the entire user study (\Cref{sec:quant:instrument}). After completing the Post-Survey, participants were sent a final email that invited them to participate in an optional \textbf{Exit Interview} (\Cref{sec:userstudy:interview}).

\subsection{Interview Procedure}
\label{sec:userstudy:interview}
For participants who indicated their desire to be interviewed, an audio-recorded Zoom interview of 20--30 minutes was conducted by a single author. The interview focused on assessing dimensions of participants' experiences that were not evident from the surveys. We began with questions about participants' \emph{motivations and routines}, which led into questions assessing the \emph{constraint page}'s alignment with their decision-making process. Next, we asked participants about the \emph{schedule page}, including how the recommended schedules factored into their decision-making and how it impacted the outcomes of their driving. Further questions focused on the \emph{earnings estimates}, including perceptions of their accuracy and whether participants would've preferred another condition. Then, we asked participants to recall a specific day of interaction in terms of how the tool affected their behaviour for that day and for the following day. Finally, participants were asked for their \emph{overall thoughts} on the tool. The full interview script is shown in \Cref{sec:app:interview}. 

\section{Quantitative Analysis}
\label{sec:quant}
Among the 51 participants in the study, 25 (49\%) were from Los Angeles, 10 (19.6\%) were from New York, 8 (15.7\%) were from Chicago, and 8 (15.7\%) were from Houston; 4 (7.8\%) were aged 18--24, 15 (29.4\%) were aged 25--34, 22 (43.13\%) were aged 35--44, 8 (15.7\%) were aged 45--54, and 2 (3.9\%) were aged over 55; 34 (66.7\%) were male, 15 (29.4\%) were female, and 1 (2\%) was non-binary; 5 (9.8\%) had a graduate degree, another 16 (31.4\%) had an undergraduate degree, another 11 (21.6\%) had a professional degree, and another 19 (37.3\%) had a high school degree.

\begin{figure*}[ht]
    \centering
    \includegraphics[trim={0 11.25cm 0 0},clip,width=\linewidth]{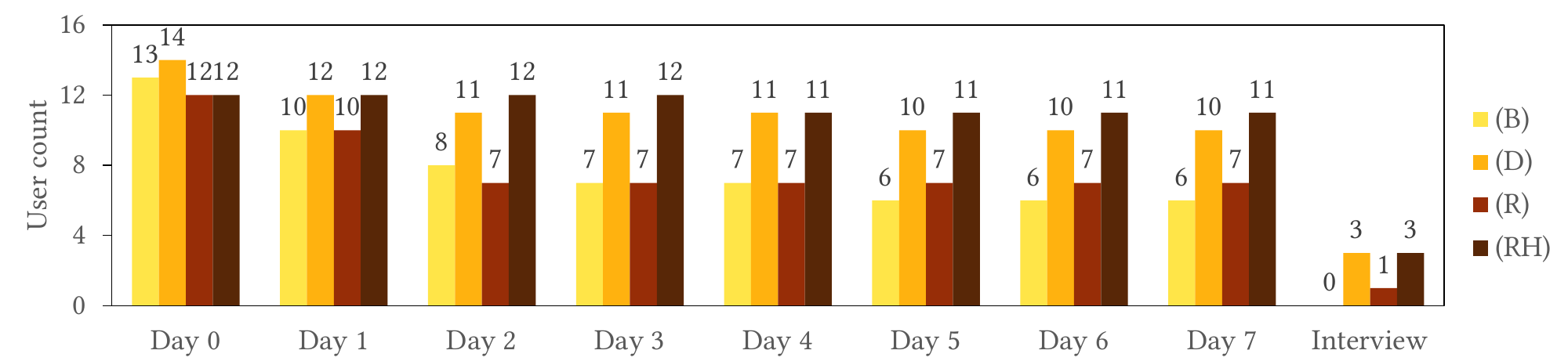}
    \caption{Retention statistics for the longitudinal user study, decomposed by design condition. Each day is labelled with the number of participants who completed all study activities for that day. Note the higher retention for Conditions~(D) and (RH).}
    \label{fig:retention}
    \Description{Bar graph showing the number of participants from Day 0 to Day 7, followed by the Interview. On Day 0, the four conditions respectively had 13, 14, 12, and 12 participants. Over the following week, retention for Conditions (D) and (RH) remained high, but retention dropped substantially for Conditions (B) and (R). By Day 7, the four conditions respectively had 6, 10, 7, and 11 participants. The Interview had 0, 3, 1, and 3 participants for the four conditions.}
\end{figure*}

Out of these 51 participants, 44 completed at least one day of interaction with the tool, and 34 completed all 7 days of interaction. Starting from Day 0, Day 7 was reached by 6 (46\%) of the base Condition~(B) participants; 10 (71\%) of the daily estimate Condition~(D) participants; 7 (58\%) of the ranged estimate Condition~(R) participants; and 11 (92\%) of the ranged and hedged Condition~(RH) participants. We show full retention statistics in \Cref{fig:retention}. 

In the following sections, we first describe the metrics that we used to measure the participants' trust and reliance (\Cref{sec:quant:instrument}). Then, we analyse our findings from statistical models for these metrics, specifically those relating to longitudinal effects (\Cref{sec:quant:longitudinal}), and the effects of specific conditions (\Cref{sec:quant:condition}).

\subsection{Metrics of Trust and Reliance}
\label{sec:quant:instrument}
We measured the \textbf{trust} of participants using self-reported measures, following common practice \cite{Kohn2021}. We used two widely-used instruments for self-reported trust: the Human-Computer Trust Questionnaire (HCT) \cite{Madsen2000} and the Trust in Automation Scale (TiA) \cite{Jian2000}. HCT measures 5 facets of trust using 5 questions each, while TiA measures both trust and distrust with 12 questions. On \textbf{Day~0} (pre-interaction), we included 5 items, one taken from each of the HCT's 5 facets of trust, in the Pre-Survey (\Cref{sec:app:survey:2}). On \textbf{Days~1--7} (during interaction), we included 3 items taken from 3 of the HCT's 5 facets of trust, in the End-of-Day Survey (\Cref{sec:app:survey:3}). On \textbf{Day~7} (post-interaction), we also included 5 items from the TiA in the Post-Survey (\Cref{sec:app:survey:4}), with 3 measuring trust and 2 measuring distrust. All of these questions were presented to participants as 5-point Likert-type scales \cite{Chyung2017,Chyung2018}. From each survey, we computed an overall trust score by first inverting items measuring distrust, if any, and then averaging the Likert-scale responses.

We measured the \textbf{reliance} of participants, i.e. the external behavioural expression of trust, using both self-reported measures (End-of-Day Survey, Question 8; \Cref{sec:app:survey:3}) and their actual behaviour of discontinuing study participation. Specifically, we computed it as an ordinal variable with three levels: 1, if the participant indicated in the End-of-Day Survey that they intended to rely on the tool \emph{more} tomorrow; 0, if the participant indicated that they intended to rely on the tool \emph{about the same} tomorrow; and -1, if the participant indicated that they intended to rely on the tool \emph{less} tomorrow, or did not complete the next day's study activities.

We use this notation to describe our statistical models: 
\begin{itemize}    
    \item \texttt{pre\_trust\_score}: The \textbf{Day~0} (Pre-Survey) trust score.
    
    \item \texttt{trust\_score}: The \textbf{current day}'s (End-of-Day) trust score.
    
    \item \texttt{reliance}: The \textbf{current day}'s (End-of-Day) reliance score.
    
    \item \texttt{post\_trust\_score}: The \textbf{Day~7} (Post-Survey) trust score.
    
    \item \texttt{day}: The day of interaction with the schedule recommendation tool (1--7).
    
    \item \texttt{user\_id}: A randomly-assigned UUID for each participant, used as random effects.
    
    \item \texttt{condition}: The participant's schedule page condition.
    
    \item \texttt{estimate\_accurate}: A binary indicator of whether the participant perceived their earnings to be about the same as \emph{the tool's estimate} (End-of-Day Survey, Question 4).
\end{itemize}

\subsection{RQ1: Longitudinal Effects}
\label{sec:quant:longitudinal}

\subsubsection{Effects on Trust}
\label{sec:quant:longitudinal:trust}
\begin{figure*}[ht]
    \centering
    \includegraphics[width=0.8\linewidth]{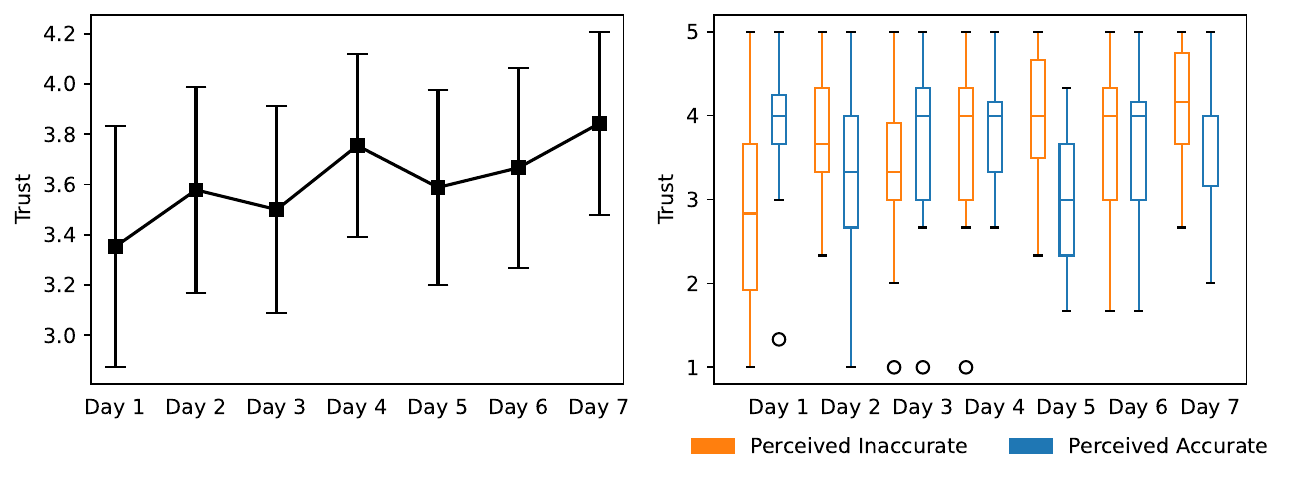}
    \caption{(L) Means and 95\% CIs of trust scores for the schedule recommendation tool on Days 1--7 among retained participants. Full statistics are shown in \Cref{tab:daily-stats} in \Cref{sec:app:quant:daily-stats}. (R) Boxplots of trust scores on Days 1--7, decomposed by perceived accuracy.}
    \label{fig:daily-trust}
    \Description{On the left, a line plot with seven points increases from 3.424 on Day 1 to 3.843 on Day 7. The general trend is upwards but not monotonic. 95\% confidence intervals for each of the points are overlapping. On the right, a boxplot shows two boxes for each day: one red (perceived inaccurate) and one green (perceived accurate). For Day 1, the IQR of the red box is below the IQR of the green box. The pattern is less clear for subsequent days: the means of the red boxes are sometimes higher and sometimes lower, but the IQRs are all overlapping.}
\end{figure*}

Participants reported a moderately high level of trust in the schedule recommendation tool ($\mu = 3.631, \sigma^2 = 0.936$). To begin, we analysed how participants' trust in the schedule recommendation tool changed over time. For the 33 retained participants who completed all 7 days of interaction, \Cref{fig:daily-trust} (left) shows an upward trend in the mean trust score. To address \Cref{rsq:accuracy}, we then grouped each day's trust scores based on whether or not participants perceived the tool's estimates as being accurate. \Cref{fig:daily-trust} (right) shows that, on Day~1, perceived accuracy was positively correlated with trust; the interquartile ranges of the trust scores did not overlap between the two groups. This effect was less clear for Days 2--7, where trust scores for the two groups overlapped more extensively.

To further explore the longitudinal effects of perceived accuracy on trust, we fitted a linear mixed model (LMM) for \texttt{trust\_score} using the R packages lme4 1.1-35.5 \cite{Bates2015} and lmerTest 3.1-3 \cite{Kuznetsova2017}. In this model, these longitudinal effects were modelled by the inclusion of the \texttt{day}, perceived accuracy (\texttt{estimate\_accurate}), and their interaction as independent variables. We also included the \texttt{pre\_trust\_score} to adjust for participants' baseline level of trust in the tool (not on the same scale), and participant IDs as random effects to account for individual variance.
\begin{align*}
    \texttt{trust\_score} \sim\ &\texttt{pre\_trust\_score} + \texttt{day} * \\
    &\texttt{estimate\_accurate} + (1 \mid \texttt{user\_id})
\end{align*}

Our model (\Cref{tab:trust-model} in \Cref{sec:app:quant:lmms}) found that participants' pre-interaction trust was significantly and positively correlated with daily trust (\texttt{pre\_trust\_score}: $\beta = 0.471, SE = 0.119, p = 0.00029$). Therefore, \textbf{participants' baseline trust persisted throughout their interactions with the tool}. Consistent with \Cref{fig:daily-trust}, their trust also increased significantly with each passing day (\texttt{day}: $\beta = 0.130, SE = 0.027, p < 0.00001$). Also consistent with \Cref{fig:daily-trust}, perceived accuracy was significantly and positively correlated with trust (\texttt{estimate\_accurate}: $\beta = 0.415, SE = 0.167, p = 0.01357$), but it had less of an impact on trust with each passing day of the user study (\texttt{day:estimate\_accurate}: $\beta = -0.121, SE = 0.037, p = 0.00126$). This suggests that, \textbf{by the end of the user study, participants' trust was based less explicitly on perceived accuracy}.

\subsubsection{Effects on Reliance}
\label{sec:quant:longitudinal:reliance}
Most participants indicated their desire to maintain their level of reliance on the schedule recommendation tool, corresponding to a reliance score of 0 ($\mu = 0.038, \sigma^2 = 0.643$). Trust and reliance were not strongly correlated ($R^2 = 0.099$); some participants consistently expressed high reliance but also lower trust. The mean reliance score appeared to decrease over time, with the mean being lowest on Day~4, but we could discern no clear dependence on perceived accuracy (\Cref{fig:daily-reliance} in \Cref{sec:app:quant:daily-stats}). To clarify the nature of these longitudinal effects, we fitted another LMM using \texttt{lme4} and \texttt{lmerTest}. This model was similar to the model for trust, except the \texttt{reliance} score was the dependent variable, and we included the \texttt{trust\_score} as an independent variable:
\begin{align*}
    \texttt{reliance} \sim\ &\texttt{pre\_trust\_score} + \texttt{day} * (\texttt{estimate\_accurate} \\
    &+ \texttt{trust\_score}) + (1 \mid \texttt{user\_id})
\end{align*}

Our model (\Cref{tab:reliance-model} in \Cref{sec:app:quant:lmms}) did not find significant effects for either the tool's perceived accuracy (\texttt{estimate\_accurate}) or the \texttt{pre\_trust\_score}. However, two effects were significant: a negative effect from the \texttt{day}, supporting our initial observation ($\beta = -0.220, SE = 0.089, p = 0.01444$), and a positive effect from the \texttt{day:trust\_score} interaction ($\beta = 0.058, SE = 0.024, p = 0.01640$). The latter suggests that reliance depended on perceived accuracy indirectly through trust. \textbf{Participants who trusted the tool more were more likely to continue relying on it}; this effect strengthened over interactions even as overall reliance weakened.

\subsection{RQ2: Effects of Conditions}
\label{sec:quant:condition}

\subsubsection{Pre-Interaction Trust}
\label{sec:quant:condition:pre}
Next, we analysed the effects of the tool's design conditions on trust and reliance, starting with pre-interaction trust. Conditions~(B)/(D)/(R)/(RH) had mean pre-interaction trust scores of 3.338, 3.629, 4.183, and 3.367; Condition~(B) had the lowest, and Condition~(R) had the highest. 

To verify these initial observations, we used the Python package \texttt{statsmodels} 0.14.2 \cite{Seabold2010} to fit an ordinary least squares (OLS) model for pre-interaction trust, with the \texttt{condition} as an independent variable. Relative to Condition~(B), the daily estimate Condition~(D) did not significantly differ in pre-interaction trust (contrast \texttt{(D)-(B)}: $\beta = 0.290, SE = 0.369, p = 0.43188$); neither did the ranged and hedged estimate Condition~(RH) (contrast \texttt{(RH)-(B)}: $\beta = 0.028, SE = 0.384, p = 0.94139$). Yet, Condition~(R) had significantly higher pre-interaction trust relative to Condition~(B) (contrast \texttt{(R)-(B)}: $\beta = 0.845, SE = 0.384, p = 0.02764$) and Condition~(RH) (contrast \texttt{(RH)-(R)}: $\beta = -0.817, SE = 0.391, p = 0.03685$). Therefore, \textbf{exposing uncertainty through range-based estimates initially improved participants' trust}. In \Cref{sec:app:quant:post}, we report on a similar analysis for post-interaction trust.

\subsubsection{Longitudinal Trust and Reliance}
\label{sec:quant:condition:longitudinal}
\begin{figure*}[ht]
    \centering
    \includegraphics[width=0.8\linewidth]{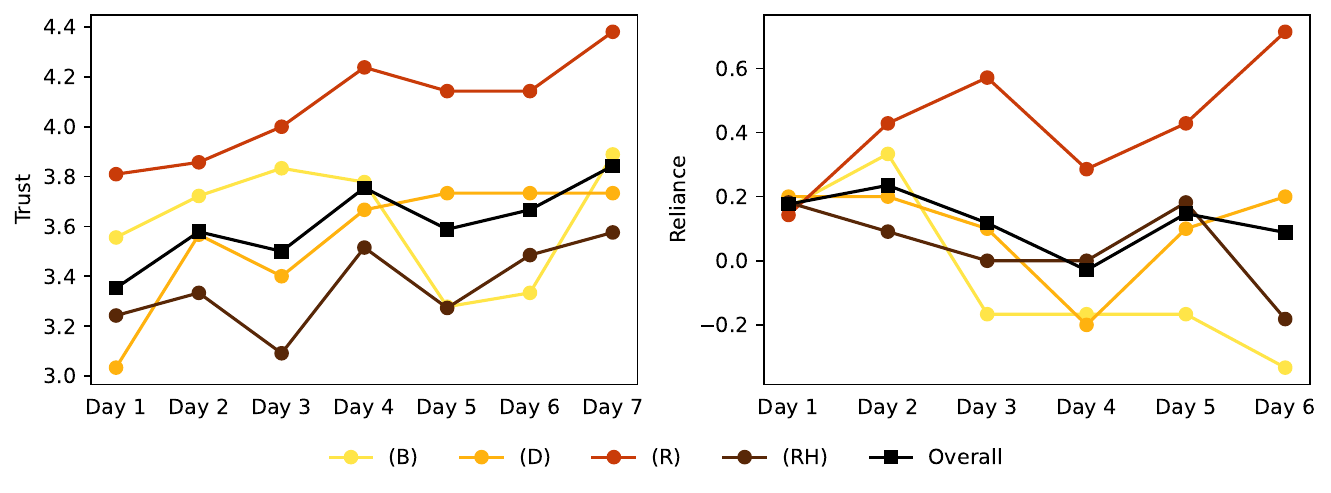}
    \caption{Mean trust (L) and reliance (R) scores for the schedule recommendation tool among retained participants, decomposed by condition. Note the higher mean scores for Condition~(R). Full statistics are shown in \Cref{tab:daily-stats} in the appendix.}
    \label{fig:condition-metrics}
    \Description{Two line plots with five lines; the lines in the left plot each have seven points representing mean trust scores on Days 1--7, while the lines in the right plot each have six points representing mean reliance scores on Days 1--6. The lines represent the overall mean and Conditions (B), (D), (R), and (RH). Between the lines, Condition (R) is consistently higher than all others; the others are all close to each other, but Condition (RH) tends to have marginally lower trust and Condition (B) marginally lower reliance.}
\end{figure*}

Lastly, we assessed the longitudinal effects of the schedule page design condition on trust and reliance. In \Cref{fig:condition-metrics}, we show the mean trust and reliance scores of participants in each of the four conditions. The means of the conditions were in most cases similar to the overall mean, with two exceptions: (1) the mean trust and reliance scores for Condition~(R) were the highest of all conditions and showed a generally increasing trend; and (2) the trust scores for Condition~(D) were lower on Days~5 and 6. To validate these trends, we added the \texttt{condition} and its interaction with the \texttt{day} as independent variables to the LMMs that we fitted in \Cref{sec:quant:longitudinal}:
\begin{align*}
    \texttt{trust\_score} \sim\ & \texttt{pre\_trust\_score} + \texttt{day} * (\texttt{estimate\_accurate} \\
    &+ \texttt{condition}) + (1 \mid \texttt{user\_id}) \\
    \texttt{reliance} \sim\ & \texttt{pre\_trust\_score} + \texttt{day} * (\texttt{estimate\_accurate} \\
    &+ \texttt{trust\_score} + \texttt{condition}) + (1 \mid \texttt{user\_id})
\end{align*}

Condition~(D) decomposed estimated earnings on a daily basis, thus providing information irrelevant to uncertainty. Our models (\Cref{tab:trust-model,tab:reliance-model} in \Cref{sec:app:quant:lmms}) indicate that this did not significantly improve trust (\texttt{condition=(D)}: $\beta = -0.494, SE = 0.341, p = 0.15139$) or reliance (\texttt{condition=(D)}: $\beta = 0.028, SE = 0.294, p = 0.92461$) over the base Condition~(B). While our trust model did find a significant, positive longitudinal effect for Condition~(D) ($\beta = 0.111, SE = 0.053, p = 0.03548$), \Cref{fig:condition-metrics} suggests that this effect does not represent a practically significant trend.

Condition~(R) displayed uncertainty in predicted earnings using ranges of pessimistic and optimistic earnings. Again, our models did not find significant marginal effects for Condition~(R) over Condition~(B) in trust (\texttt{condition=(R)}: $\beta = -0.169, SE = 0.360, p = 0.64038$) or reliance (\texttt{condition=(R)}: $\beta = -0.145, SE = 0.308, p = 0.63782$). However, Condition~(R) had nearly significant longitudinal effects for trust (\texttt{day:condition=(R)}: $\beta = 0.103, SE = 0.057, p = 0.06976$) and also reliance (\texttt{day:condition=(R)}: $\beta = 0.137, SE = 0.072, p = 0.05958$). This aligns with our observations based on the mean scores in \Cref{fig:condition-metrics} as well as our findings for pre-interaction trust (\Cref{sec:quant:condition:pre}). Therefore, despite exposing uncertainty in the tool's earnings estimates, \textbf{the ranges of Condition~(R) improved participants' initial trust and then led them to maintain their trust and reliance} over daily interactions.

Condition~(RH) added lexical hedging to the range-based earnings estimates in Condition~(R). This condition was not significantly different from Condition~(B) in marginal or longitudinal effects on trust and reliance. On Day 6, participants in Condition~(RH) reported significantly lower reliance than participants in Condition~(R) (\Cref{fig:daily-reliance}; $\mu = 0.714, -0.182; 95\%\ \textrm{CIs} = (0.240, 1.188)$, $(-0.649, 0.286)$), the only such significant pairwise difference on a daily basis (see \Cref{tab:daily-stats} in \Cref{sec:app:quant:daily-stats}). Combined with the pre-interaction trust of Condition~(RH) being significantly lower than Condition~(R) (\Cref{sec:quant:condition:pre}), we conclude that \textbf{the addition of lexical hedging in Condition~(RH) reversed the gains in trust and reliance from Condition~(R)'s range-based uncertainty}. 

\section{Qualitative Analysis}
\label{sec:qual}
Overall, 7 participants completed the exit interview after they completed all seven days of the longitudinal user study. Three of these were from Condition~(D), one was from Condition~(R), and three were from Condition~(RH):
\begin{itemize}
    \item \textbf{P1}: A 47-year-old male with a high school degree driving for Instacart and Lyft in Los Angeles, (D)
    \item \textbf{P2}: A 42-year-old female with a high school degree driving for Uber Eats in Los Angeles, (D)
    \item \textbf{P3}: A 29-year-old female with a graduate degree driving for Lyft in Chicago, (RH)
    \item \textbf{P4}: A 49-year-old male with an undergraduate degree driving for Lyft and Uber in Los Angeles, (RH)
    \item \textbf{P5}: A 46-year-old male with an undergraduate degree driving for DoorDash, GrubHub, and Uber Eats in Chicago, (R)
    \item \textbf{P6}: A 39-year-old male with a graduate degree driving for Lyft, Uber, and Uber Eats in Los Angeles, (D)
    \item \textbf{P7}: A 39-year-old male with a professional degree delivering for DoorDash in Los Angeles by bike, (RH)
\end{itemize}

Transcripts for the interviews were generated by Zoom. The author who conducted the interviews reviewed and corrected these transcripts, then performed structural coding \cite{Saldana2015a}. Afterwards, two authors separately used QualCoder 3.3 \cite{Curtain2023} to perform open coding \cite{Saldana2015a} and axial coding \cite{Saldana2015b}. The two authors met to reconcile their codes and construct a unified codebook. Finally, the interviewing author re-applied the updated codes. Now, we discuss our findings in relation to participants' motivations for using the tool (\Cref{sec:qual:motivation}), perceptions of its accuracy (\Cref{sec:qual:accuracy}), and perceptions of its uncertainty based on the design conditions (\Cref{sec:qual:uncertainty}).

\subsection{Motivations and Routines}
\label{sec:qual:motivation}
\textbf{Participants reported a diversity of motivations and routines for gig driving, which impacted their perceptions of the schedule recommendation tool's usefulness.} P1 and P2 viewed gig driving as a primary source of income, and thus found more value in the tool's earnings estimates: 
\textit{``[The tool was] definitely worthwhile, just because it gave me a number, a projection. [...] They definitely motivate me to keep going the next day.'' -- P1~(D)}
Meanwhile, P3--P7 used gig driving to supplement other sources of income, but P4 still viewed his earning goals as important. Unlike other participants, P7 delivered with a bicycle in his spare time. He felt that his current commitment was insufficient to want to use the tool more:  
\textit{``If I take this job to a full time, take it seriously? I would [want to use it more].'' -- P7~(RH)} 
Nevertheless, drivers found value in the tool regardless of their level of motivation. P1--P6 all reported challenges in estimating their potential earnings as a consequence of unpredictability in gig demand, pay, or location, or of gig platforms providing insufficient information. For instance:
\textit{``Uber's details that they offer to drivers through their interfaces are sorely lacking. So I'm grateful for the opportunity to interact with this tool.'' -- P6~(D)} 

\subsection{RQ1: Perceptions of Accuracy}
\label{sec:qual:accuracy}
When evaluating the tool's accuracy, participants weighed its recommendations against their own routines and intuitions. For P1, P3, P6, and P7, the tool was a reference for how well they could perform in their existing routines, rather than something to reshape their routines:
\textit{``I still would've followed my routine. [...] I was fortunate enough to at least have the tool make me a schedule based on the routine that I currently do.'' -- P3~(RH)} 
However, P5 suggested that the tool could use a question-answering approach to nudge users into altering their routines, by first understanding their activity patterns and then suggesting modifications. When the tool was inaccurate, participants reacted in different ways. P1, P2, and P4 observed that instances of the tool being inaccurate decreased their desire to comply with the tool's recommendations: 
\textit{``If I was making more than what it said, I would have done it more consistently on the schedule.'' -- P2~(D)}
P4's reactions to inaccuracies were influenced by his expectations. He was motivated on one instance by the tool's estimates exceeding his goals:
\textit{``My target's [...] \$30 an hour. Because those [estimated earnings] were consistently below \$30, [...] I wasn't motivated to study it. But when I saw the 4 to 6 am, that kind of piqued my interest.'' -- P4~(RH)} 
\textbf{Maintaining consistent perceptions of accuracy over time was important for building trust in this context}. P1, P2, and P5 indicated that the outcomes of their first one or two days of interaction impacted their willingness to follow the recommendations for the rest of the study. P4 and P5 indicated that their use of the tool would be strengthened longitudinally if they consistently perceived its predictions as being accurate: 
\textit{``Once I learned that it was accurate, and I had trust in it, and it was really helping, then I'd probably use it more and more.'' -- P4~(RH)}

\subsection{RQ2: Perceptions of Uncertainty}
\label{sec:qual:uncertainty}
Some participants recognised that the accuracy of the tool's earnings estimates would be impacted by both their own decisions (P1, P3) and other environmental factors (P6):
\textit{``It also depends, too, on the rides that I accept.'' -- P3~(RH)}
\textit{``It might be true that I might earn the forecasted average earnings. But surges can definitely make a difference.'' -- P6~(D)} 
Note that the tool's uncertainty was not exposed to P6, suggesting that this observation originated from their innate mental model. Recognising the effect of their own agency led P1 and P3, as well as P5, to adopt the tool's estimates as goals for their own earnings:
\textit{``[...] setting daily goals of how much money I would like to make [...] was definitely something that I wasn't really doing prior to doing this study or using this tool.'' -- P3~(RH)}
Also, P1 suggested that being able to compare their earnings to the tool's estimates in an hourly breakdown would be helpful for goal-setting.

All participants in Conditions~(R) and (RH) (P3--P5, P7) appreciated the presence of ranges. P5 compared the tool's range to his own experiences:
\textit{``I was always over the average. So, to me, I was kind of in my head using that as a low.'' -- P5~(R)}
For participants in the other conditions, P1 and P2 indicated that they would've preferred to have had ranges. However, P6 suggested that ranges may lead to disappointment when they are used for goal-setting:
\textit{``If they earn less than [the higher number], then they probably might feel disappointed in the tool through no fault of its own, right? If you say \$12 to \$18, and it comes in at \$14, [...] I could understand how folks might look at that as a let down.'' -- P6~(D)}
Thus, \textbf{range-based uncertainty was useful for decision-making, but needed to be calibrated against expectations}.

Both P4 and P5 struggled with the idea that the uncertainty in the tool's estimates could have originated from drivers with habits different to themselves:
\textit{``Obviously no one's ever gonna work if it's just \$4 or \$5 an hour.'' -- P4~(RH)}
This was in spite of the lexical hedging presented to P4. At least for this participant, the verbiage in our hedges may thus have failed to achieve its goal of leading him to consider potential sources of uncertainty more carefully.

\section{Discussion}
\label{sec:disc}

\subsection{Key Findings and Implications}
\label{sec:disc:findings}

\textbf{Trust in AI decision aids is built both initially and over time.} 
We found that participants' pre-interaction trust in our schedule recommendation tool significantly impacted their trust during interaction (\Cref{sec:quant:longitudinal,sec:quant:condition:pre}). This is consistent with findings in the medical domain that practitioners \cite{Burgess2023,Cai2021} and patients \cite{Mou2017a,Mou2017b} prefer to gauge their trust prior to interaction. Our interviews similarly showed that perceptions of the tool's accuracy in the first two days influenced subsequent trust (\Cref{sec:qual:accuracy}). Yet, we also found that trust and reliance increased across interactions with the tool. While perceived accuracy had diminishing impacts on trust in later stages of the user study (\Cref{sec:quant:longitudinal}), P4's experience (\Cref{sec:qual:accuracy}) shows that critical incidents where estimates differ significantly from expectations can cause catastrophic losses of trust \cite{Stanley2023}. P4 found this difficult to recover from. However, losses of trust could be mitigated prospectively by calibrating perceptions of accuracy, e.g. by emphasising that drivers' outcomes are also a function of their own decisions. This could foster \emph{appropriate reliance} by helping users decide when or when not to rely on the tool \cite{Schemmer2023}.

\textbf{Interactivity could help to maintain trust over time.} Losses of trust can also be mitigated retrospectively based on trust repair strategies \cite{Pareek2024}. Based on our qualitative findings, we hypothesise two mechanisms by which interactivity could help to maintain and repair trust. First, interactivity may enhance perceptions of control. The modes of interaction suggested by our participants, such as hourly breakdowns and question-answering, would assure users that the AI has the intent and agency to capture and learn from their preferences \cite{Pareek2024}. Second, interactivity could help users to better recall their experiences and decisions. Until the interviewer probed further, most interview participants could not recall whether their earnings significantly differed from the tool's estimates. 

\textbf{The impact of exposing uncertainty on trust in AI decision aids depends on task alignment.} 
Prior work has reached mixed conclusions on how exposing uncertainty in AI impacts trust and reliance. On similar tasks, \citet{Zhang2020b} found that confidence scores improve reliance, whereas \citet{Prabhudesai2023} found that distribution plots dampen trust and reliance. \citet{Yang2024} found that the effects of these designs depended on individual characteristics, but our results suggest another dimension: task alignment in the designs themselves. Task-aligned uncertainty representations, i.e. scalar ranges as opposed to distributions, allowed our participants to incorporate uncertainty directly into their decision-making (\Cref{sec:qual:uncertainty}), thus improving trust (\Cref{sec:quant:condition:longitudinal}). This is consistent with findings in the AI explainability literature that domain-aligned explanations are more persuasive \cite{Cau2023,Naiseh2023}. We hypothesise that task alignment also underlies the negative effect of hedging we observed (\Cref{sec:quant:condition:longitudinal}): thinking about other drivers is not helpful when drivers are trying to reason about their own outcomes (\Cref{sec:qual:uncertainty}).

\textbf{How uncertainty is exposed should be adapted to user subpopulations.} Our results did not find that a one-size-fits-all approach exists to fostering trust. Even within the same condition, participants exhibited variability in how they reacted to the outcomes of their reliance. Nevertheless, we hypothesise that it may be helpful to adapt uncertainty in predictions and recommendations to subgroups within the gig driver population. Specifically, our qualitative results point to differing perceptions of accuracy and uncertainty between highly motivated drivers (e.g. P1 and P2) and less motivated drivers (e.g. P7). If drivers received estimates from those with habits similar to themselves, this could mitigate P4 and P5's struggles with how to interpret uncertainty. A future large-scale study could help to confirm our hypothesis.

\subsection{Limitations}
\label{sec:disc:limitations}
Our work has two primary limitations. First, we cannot claim that the design of our tool was optimal for engendering trust. Our focus was on testing how designs for exposing uncertainty would impact trust and reliance. Thus, we attempted to isolate the effect of this design dimension by refining the tool through a formative pilot study (\Cref{sec:app:pilot}). Nevertheless, further improvements may have been possible. For instance, we could not provide retrospective breakdowns of participants' earnings due to data availability limitations. Thus, design choices orthogonal to the exposure of uncertainty may have impacted participants' trust and reliance. 

Second, despite our best efforts, our sample of drivers was limited. These individuals were at least aware, if not active users, of the Gridwise app, and thus they may have been more focused on their outcomes than the general gig driver population. The trust of users in an AI decision aid is contingent upon their domain knowledge \cite{Kim2023,Wysocki2023}, and --- as we demonstrate (\Cref{sec:qual:motivation}) --- the extent to which they integrate the decision aid into their existing routines. A future study aimed at a broader population of gig drivers could uncover additional insights by explicitly controlling for factors such as full-time status and driver tenure. We were also unable to reach participants who discontinued the user study. Future studies that follow up with such participants would be a valuable source of data on mechanisms of trust loss and repair in longitudinal settings. 

\subsection{Recommendations for Future Work}
\label{sec:disc:recommend}
The paucity of similar longitudinal, in-situ studies in prior work is understandable given the logistical challenges we encountered. We stress the importance of observational studies to improve domain understanding as a basis for longitudinal interventional studies. Our pilot interviews helped us to design a tool that was task-aligned, which led participants to find value in it over repeated interactions. Furthermore, our study design aimed to increase perceived control while reducing user burden through flexibility in the scheduling of participation; customisability of the constraints on the AI decision aid; shorter survey instruments; and incremental compensation.

\section{Conclusion}
\label{sec:conc}
In this paper, we assessed how users' trust and reliance on AI decision aids is influenced by designs that expose uncertainty. Unlike the laboratory experiments used by previous work, we did so within the real-world context of gig driving. Specifically, we conducted a longitudinal, in situ user study of an AI-based schedule recommendation tool with $n = 51$ gig drivers. These drivers' interactions with our tool impacted their actual earnings. Our findings demonstrate that trust can be built by (1) maintaining perceptions of accuracy over repeated interactions and (2) displaying uncertainty in a task-aligned fashion, the latter of which points to a need for more context-specific evaluations of AI decision aids.

\begin{acks}
We thank Jenny T. Liang, Joshua Mateer, Jose M. del Alamo, Ian Yang, Chenyang Yang, Kaia Newman, Stephanie Milani, Niklas K\"uhl, and Nadia Nahar for their feedback and discussion on the study's instruments, protocol, analysis, and manuscript. Also, we thank Linda Moreci for assistance with participant compensation, and Akshara Khare and Anagha Ravi Shankara for contributions to the recommendation tool's implementation. This work was conducted in collaboration with Gridwise; we thank Kunal Prajapati, Karan Moudgil, Ryan Green, and Brian Finnamore. We were supported by NSF Grant CNS-1801316, a research grant from Mobility21, and the Tang Family Endowed Innovation Fund.
\end{acks}

\newpage
\bibliographystyle{ACM-Reference-Format}
\bibliography{ref}

\newpage
\appendix
\section{Survey Questions}
\label{sec:app:survey}

\subsection{Survey 1 --- Intake Survey}
\label{sec:app:survey:1}
\begin{enumerate}[1.]
    \item Please tell us which of the following regions you primarily drive in.
    \begin{itemize}
        \item Los Angeles
        \item New York
        \item Chicago
        \item Houston
    \end{itemize}

    \item Please tell us which of the following services you currently drive for. 
    \begin{itemize}
        \item DoorDash
        \item Grubhub
        \item Instacart
        \item Lyft
        \item Uber
        \item Uber Eats
        \item Other
    \end{itemize}
    
    \item In 2 or 3 sentences, please tell us what you like most about driving for ridesharing and/or delivery services.
    
    \item In 2 or 3 sentences, please tell us what you like least about driving for ridesharing and/or delivery services.
\end{enumerate}
    
\textit{How much do you agree or disagree with the following statements?}
    				
\begin{enumerate}[1.]
\setcounter{enumi}{4}
    \item When I drive, it's important to me that I make some minimum amount of money.
    \begin{itemize}
        \item Strongly disagree
        \item Disagree
        \item Not sure
        \item Agree
        \item Strongly agree
    \end{itemize}
    				
    \item When I drive, I have an accurate sense of how much money I will make.
    \begin{itemize}
        \item Strongly disagree
        \item Disagree
        \item Not sure
        \item Agree
        \item Strongly agree
    \end{itemize}
    				
    \item When I don't earn the amount that I expect to from driving, it causes difficulties for me.
    \begin{itemize}
        \item Strongly disagree
        \item Disagree
        \item Not sure
        \item Agree
        \item Strongly agree
    \end{itemize}
    				
    \item I try to stick to a regular routine for times and places to drive.
    \begin{itemize}
        \item Strongly disagree
        \item Disagree
        \item Not sure
        \item Agree
        \item Strongly agree
    \end{itemize}

    \item I am happy with how I currently decide when and where to drive.
    \begin{itemize}
        \item Strongly disagree
        \item Disagree
        \item Not sure
        \item Agree
        \item Strongly agree
    \end{itemize}
\end{enumerate}
    
\begin{enumerate}[1.]
\setcounter{enumi}{9}
    \item Please tell us your age.
    
    \item Please tell us what gender you identify as.
    \begin{itemize}
        \item Male
        \item Female
        \item Non-binary
        \item Prefer not to disclose
        \item Other 
    \end{itemize}
    
    \item Please tell us your highest education level.
    \begin{itemize}
        \item Less than high school
        \item High school
        \item Some two-year professional degree
        \item Some undergraduate degree
        \item Some graduate degree (MS, PhD, JD, or MD)
    \end{itemize}

\end{enumerate}

\subsection{Survey 2 --- Pre-Survey}
\label{sec:app:survey:2}
\textit{How much do you agree or disagree with the following statements?}
    				
\begin{enumerate}[1.]
    \item I understand how the tool used my answers to generate this recommended schedule.
    \begin{itemize}
        \item Strongly disagree
        \item Disagree
        \item Not sure
        \item Agree
        \item Strongly agree
    \end{itemize}
    \textit{This item is based on item U2 from the Perceived Understandability questions of the HCT \cite{Madsen2000}, ``I understand how the system will assist me with decisions I have to make.'' As the constraint page is intended to encapsulate the user's decision-making process, we consider the generation of the recommended schedule to be how the tool assists the user with their decisions.}
				
    \item I feel that I can rely on the tool to produce recommendations which accommodate the things that matter most to me.
    \begin{itemize}
        \item Strongly disagree
        \item Disagree
        \item Not sure
        \item Agree
        \item Strongly agree
    \end{itemize}
    \textit{This item is based on item R4 from the Perceived Reliability questions of the HCT \cite{Madsen2000}, ``I can rely on the system to function properly.'' We consider the tool to be properly functioning if its recommendations account for the user's goals and preferences.}
    				
    \item I feel that the driving times recommended by the tool are as good as what an experienced driver would recommend to me.
    \begin{itemize}
        \item Strongly disagree
        \item Disagree
        \item Not sure
        \item Agree
        \item Strongly agree
    \end{itemize}
    \textit{This item is based on item T3 from the Perceived Technical Competence questions of the HCT \cite{Madsen2000}, ``The advice the system produces is as good as that which a highly competent person could produce.'' Our tool's advice is its recommended schedule, and to our participants a competent individual would be an experienced driver.}
    				
    \item I feel that the times suggested by the tool are good even if I don't know for certain that they will maximise my earnings / minimise my hours \textit{[depending on the constraints selected]}.
    \begin{itemize}
        \item Strongly disagree
        \item Disagree
        \item Not sure
        \item Agree
        \item Strongly agree
    \end{itemize}
    \textit{This item is based on item F1 from the Faith questions of the HCT \cite{Madsen2000}, ``I believe advice from the system even when I don't know for certain that it is correct.'' Again, our tool's advice is its recommended schedule of driving times. Since we cannot apply a clear notion of correctness to continuous estimates of earnings, we reworded this question to focus on alignment with the user's objectives.}
    				
    \item I would like to use the tool to decide my driving hours in the future.
    \begin{itemize}
        \item Strongly disagree
        \item Disagree
        \item Not sure
        \item Agree
        \item Strongly agree
    \end{itemize}
    \textit{This item is based on item P4 from the Personal Attachment questions of the HCT \cite{Madsen2000}, ``I like using the system for decision making.'' We reworded it to better assess participants' level of intended future reliance on the tool.}
\end{enumerate}

\subsection{Survey 3 --- End-of-Day Survey}
\label{sec:app:survey:3}
\begin{enumerate}[1.]
    \item How often did you follow the times in the recommended schedule today?
    \begin{itemize}
        \item I did not follow the recommendations at all
        \item I followed the recommendations for one hour during the day
        \item I followed the recommendations for two or three hours during the day
        \item I followed the recommendations for four or more hours during the day
    \end{itemize}

    \item How satisfied do you feel you are with your earnings from today?
    \begin{itemize}
        \item Very dissatisfied
        \item Somewhat dissatisfied
        \item Neither satisfied nor dissatisfied
        \item Somewhat satisfied
        \item Very satisfied
    \end{itemize}

    \item As far as you remember, how did your earnings today compare to your expectations?
    \begin{itemize}
        \item Lower
        \item About the same
        \item Higher
        \item Not sure
    \end{itemize}

    \item As far as you remember, how did your earnings today compare to the tool's estimate?
    \begin{itemize}
        \item Lower
        \item About the same
        \item Higher
        \item Not sure
    \end{itemize}
\end{enumerate}

\textit{How much do you agree or disagree with the following statements?}

\begin{enumerate}[1.]
\setcounter{enumi}{4}
    \item I felt that the recommended schedule provided by the tool was easy to follow.
    \begin{itemize}
        \item Strongly disagree
        \item Disagree
        \item Not sure
        \item Agree
        \item Strongly agree
    \end{itemize}
    \textit{This item is based on item U4 of the Perceived Understandability questions of the HCT \cite{Madsen2000}, ``It is easy to follow what the system does.'' Instead of asking the user about the tool's operation generally, we focused the question on the interpretability of its recommended schedule for that day.}

    \item I felt that the recommended schedule provided all of the information that I needed to decide when to drive.
    \begin{itemize}
        \item Strongly disagree
        \item Disagree
        \item Not sure
        \item Agree
        \item Strongly agree
    \end{itemize}
    \textit{This item is based on item R1 of the Perceived Reliability questions of the HCT \cite{Madsen2000}, ``The system always provides the advice I require to make my decision.'' Again, our tool's advice is its recommended schedule. We focused the question on the user's decisions for that particular day.}

    \item When I was unsure of when to drive today, I followed the recommended schedule.
    \begin{itemize}
        \item Strongly disagree
        \item Disagree
        \item Not sure
        \item Agree
        \item Strongly agree
    \end{itemize}
    \textit{This item is based on item F2 from the Faith questions of the HCT \cite{Madsen2000}, ``When I am uncertain about a decision I believe the system rather than myself.'' We focused the question on the user's decisions for that particular day, and reworded ``believe'' to ``follow'' to assess compliance more clearly.}

    \textit{We left out questions based on Perceived Technical Competence and Personal Attachment for length.}
\end{enumerate}

\begin{enumerate}[1.]
\setcounter{enumi}{7}
    \item Which of the following statements do you agree with most?
    \begin{itemize}
        \item I intend to rely on the tool less tomorrow than I did today
        \item I intend to rely on the tool about the same tomorrow as I did today
        \item I intend to rely on the tool more tomorrow than I did today
        \item I intend to pause my interaction with the tool for one day tomorrow
    \end{itemize}
\end{enumerate}

\subsection{Survey 4 --- Post-Survey}
\label{sec:app:survey:4}
\textit{How much do you agree or disagree with the following statements?}
    				
\begin{enumerate}[1.]
    \item I feel that I have become familiar with how to use the tool.
    \begin{itemize}
        \item Strongly disagree
        \item Disagree
        \item Not sure
        \item Agree
        \item Strongly agree
    \end{itemize}
    \textit{This item is based on item 12 of the TiA \cite{Jian2000}, ``I am familiar with the system.'' We reworded the question in light of the fact that users did not have any existing experience with using the tool before the study.}

    \item When I am using a navigation app that suggests routes to me, I feel like I would want to follow the suggestions more if the app asked me questions about my preferences (like this tool did) before giving its suggestions.
    \begin{itemize}
        \item Strongly disagree
        \item Disagree
        \item Not sure
        \item Agree
        \item Strongly agree
    \end{itemize}
    \textit{This item is an original question that prompts the participant to consider their interactions with other types of recommendation systems. It assesses the extent to which participants would appreciate granular controls based on their preferences in such systems.}
\end{enumerate}

\begin{enumerate}[1.]
\setcounter{enumi}{2}
    \item Were there questions that you wanted the tool to ask you that it didn't? If so, please tell us about them in 2 or 3 sentences.
\end{enumerate}

\textit{How much do you agree or disagree with the following statements?}
\begin{enumerate}[1.]
\setcounter{enumi}{3}
    \item I felt that I was able to trust the schedules recommended by the tool.
    \begin{itemize}
        \item Strongly disagree
        \item Disagree
        \item Not sure
        \item Agree
        \item Strongly agree
    \end{itemize}
    \textit{This item is based on item 11 of the TiA \cite{Jian2000}, ``I can trust the system.'' We focused the scope of this question on the output of the tool, the recommended schedule, rather than the tool as a whole.}
    				
    \item I felt that I was able to depend on the schedules recommended by the tool for deciding when to drive.
    \begin{itemize}
        \item Strongly disagree
        \item Disagree
        \item Not sure
        \item Agree
        \item Strongly agree
    \end{itemize}
    \textit{This item is based on item 9 of the TiA \cite{Jian2000}, ``The system is dependable.'' Again, we focused the scope of this question on the output of the tool, the recommended schedule.}

    \item When I am using a navigation app that suggests routes to me, I feel like I would want to follow the suggestions more if the app gave me information about the minimum and maximum possible time of the trip (similar to what this tool did).
    \begin{itemize}
        \item Strongly disagree
        \item Disagree
        \item Not sure
        \item Agree
        \item Strongly agree
    \end{itemize}
    \textit{This item is an original question that prompts the participant to consider their interactions with other types of recommendation systems. It assesses the extent to which participants would appreciate increased exposure of uncertainty through range-based estimates in such systems.}
    				
    \item I felt that the recommended schedule was misleading.
    \begin{itemize}
        \item Strongly disagree
        \item Disagree
        \item Not sure
        \item Agree
        \item Strongly agree
    \end{itemize}
    \textit{This item is based on item 1 of the TiA \cite{Jian2000}, ``The system is deceptive.'' In addition to focusing the scope of this question on the output of the tool, we also reworded ``deceptive'' to ``misleading'' to capture the broader possibility of the tool being perceived as unintentionally providing incorrect information.}
    				
    \item I felt that the recommended schedule harmed my earnings.
    \begin{itemize}
        \item Strongly disagree
        \item Disagree
        \item Not sure
        \item Agree
        \item Strongly agree
    \end{itemize}
    \textit{This item is based on item 5 of the TiA \cite{Jian2000}, ``The system's actions will have a harmful or injurious outcome.'' In our context, the outcome for the user is their earnings from driving while following the recommended schedule. We reworded the question to assess the outcome retrospectively.}
\end{enumerate}

\begin{enumerate}[1.]
\setcounter{enumi}{8}
    \item If there were any, please identify some of the driving times recommended by the tool that did not align with your expectations.
    \begin{itemize}
        \item When was the time?
        
        \item In 1 or 2 sentences, why did it not align with your expectations?
    \end{itemize}

    \item Do you have any other questions or comments regarding this tool that you would like to share with us?
\end{enumerate}

\section{Pilot Interviews}
\label{sec:app:pilot}
To assess the utility of the AI-based schedule recommendation tool design introduced in \Cref{sec:tool}, we created a prototype based on it. We then conducted a series of interviews to understand gig drivers' needs and how well the prototype aligned with them. The interview methodology was approved by our Institutional Review Board (IRB).

\subsection{Tool Design}
\label{sec:app:pilot:tool}
Following common practice in UX design \cite{Feng2023}, we used Figma \cite{Figma} to create the prototype design (\Cref{sec:app:figma}). 

\begin{figure*}[ht]
    \centering
    \includegraphics[width=0.8\linewidth]{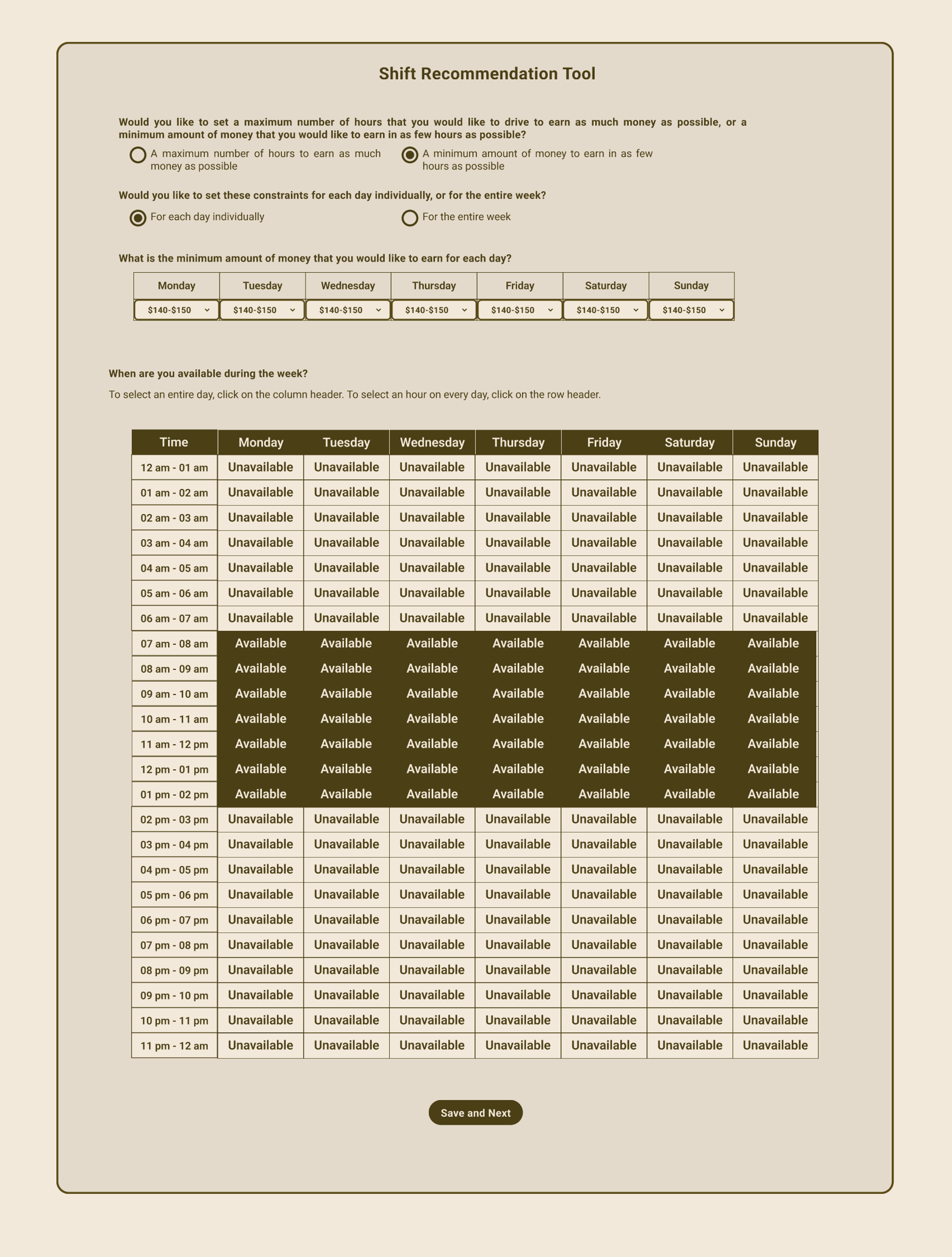}
    \caption{Screenshot of Figma prototype for the constraint page. On behalf of participants, the interviewer clicked on radio buttons, input fields, and table cells to set their constraints. For technical reasons, input fields for the numerical constraints were implemented as dropdown menus.}
    \label{fig:figma:constraint}
    \Description{Screenshot of a webpage titled "Shift Recommendation Tool". The constraint page is shown, with various inputs being set to mock values. In response to "Would you like to set a maximum number of hours that you would like to drive to earn as much money as possible, or a minimum amount of money that you would like to earn in as few hours as possible?", the latter is selected. In response to "Would you like to set these constraints for each day individually, or for the entire week?", the former is selected. In response to "What is the minimum amount of money you would like to earn for each day?", \$140-\$150 is entered for each day of the week. In response to "When are you available during the week?", time slots corresponding to 7 am to 2 pm on every day of the week are selected (as indicated by dark cells in a tabular schedule).}
\end{figure*}

The \textbf{constraint page}'s design (\Cref{fig:figma:constraint}) was largely similar to the final design. However, the prototype had a monotone colour scheme and had instructions that were worded less clearly, which were improved based on feedback from the pilot (\Cref{sec:app:pilot:results}). Additionally, due to technical limitations, dropdowns were used in place of text boxes. However, we consider the impact of these limitations to be minor since the interviewer controlled the page. 

\begin{figure*}[ht]
    \centering
    \includegraphics[width=0.8\linewidth]{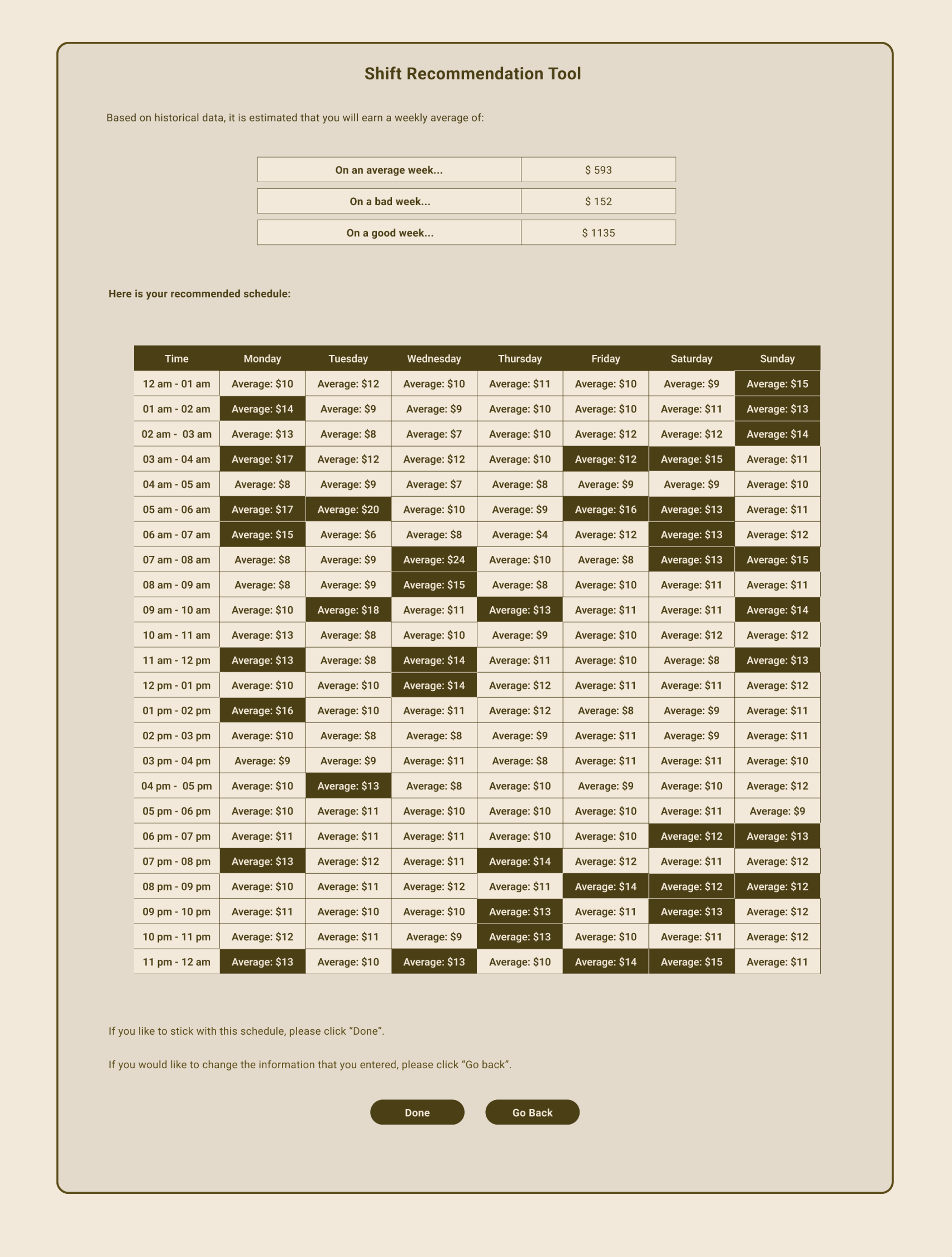}
    \caption{Screenshot of Figma prototype for the schedule page. A summary of mean, pessimistic, and optimistic weekly earnings is shown at the top of the page, followed by an hourly schedule where recommended cells are highlighted in darker colours. For technical reasons, this was shown as a static page not depending on previously-entered constraints.}
    \label{fig:figma:schedule}
    \Description{Screenshot of a webpage titled "Shift Recommendation Tool". The schedule page is shown with mock estimates. The top of the schedule page reads, "Based on historical data, it is estimated that you will earn a weekly average of". This is followed by a table that reads, "On an average week: \$593", "On a bad week: \$152", and "On a good week: \$1135". Further below, a tabular schedule is shown for each day of the week. Each cell reads, "Average", followed by the estimated hourly earning in bold. Recommended time slots are indicated by dark cells, and non-recommended time slots are indicated by light cells.}
\end{figure*}

The \textbf{schedule page}'s design (\Cref{fig:figma:schedule}) was kept static for the pilot to gather more uniform feedback from participants. Like the constraint page, the prototype differed from the final design in its colour scheme and clarity of wording. The prototype was most similar to Condition~(R) (\Cref{sec:tool:conditions}) in that it displayed a range consisting of mean (``On an average week...''), pessimistic (``On a bad week...''), and optimistic (``On a good week...'') weekly earnings above the tabular schedule. Unlike the final design of Condition~(R), however, ranges were not shown for hourly estimates; this was added based on feedback from the pilot (\Cref{sec:app:pilot:results}). 

\subsection{Methodology}
\label{sec:app:pilot:methods}
Participants began by completing a web-based consent form and a demographics survey that collected their age, gender, and education level. This form was separate from the one for the longitudinal user study (\Cref{sec:userstudy}). After completing the web form, participants were invited via email to complete 20--30-minute audio-recorded Zoom interviews conducted by a single author. Participants were compensated with a \$10 Amazon gift card.

In the first 5--10 minutes, the interview focused on \emph{formatively} understanding drivers' needs and motivations. The questions asked in this portion of the interview were the same as the Intake Survey of the longitudinal user studies (\Cref{sec:userstudy:activities}). In the last 15--20 minutes, the interview focused on \emph{evaluatively} understanding how well the tool met drivers' needs. To ensure a consistent experience, the Figma prototype was opened in the interviewer's browser and shown to participants in a screensharing session. First, on the constraint page, the participant was asked to work with the interviewer to interact with the page, entering the constraints as if they were using the tool for their actual planning. Then, on the schedule page, the participant was shown a schedule with mocked earnings estimates. Finally, the participant was asked about their overall opinions of the tool. The full interview script is shown in \Cref{sec:app:pilot:script}.

To analyse these interviews, we used the same methodology as the longitudinal interviews (\Cref{sec:qual}). 

\subsection{Interview Script}
\label{sec:app:pilot:script}

\subsubsection{Formative Questions}
\label{sec:app:pilot:script:formative}
\textit{As the interviews were semi-structured, the script below focuses on the guiding questions that we asked participants. The interviewer also probed participants further depending on their responses.}

\begin{itemize}
    \item Please tell us what you like most about driving for ridesharing and/or delivery services.
    \item Please tell us what you like least about driving for ridesharing and/or delivery services.
    \item When you drive, how important is it to you that you make some minimum amount of money daily/weekly?
    \item When you drive, do you have an accurate sense of how much money you will make?
    \item Do you try to stick to a regular routine for times and places to drive? 
    \item Are you happy with your current routines in terms of when and where you drive?
    \item Would getting recommendations for times to drive would be helpful to you?
\end{itemize}

\subsubsection{Evaluative Questions}
\label{sec:app:pilot:script:evaluative}
We will show you a tool that can suggest personalised driving schedules. The tool will ask you some questions about your availability and preferences, as well as revenue targets that you might have. Different people might want to use the tool differently. However, we expect a typical user to use it as follows. 

First, they would fill in some information about when they are available during the week, along with either how long they want to work or how much they want to make. The tool will then suggest a recommended schedule for the entire week. As they return to the tool every day to plan out their schedules, users will have the opportunity to interact with the tool, tweaking their availability and possible revenue targets to see how the recommendations change.

\textit{The interviewer begins screensharing the constraint page prototype.}

\begin{itemize}
    \item Here's the initial page of the tool that lets you specify your availability and goals.
    \item Do you believe you understand what is being shown on this page?
    \item Do you feel that this tool is asking you the right questions about your availability and goals?
    \item Are there other important questions that you wish the tool would ask?
    \item Think about your upcoming week. Using this screen, please tell us what information you think you would want to enter to get a useful recommendation. We will click on the page for you.
\end{itemize}

\textit{The interviewer switches to the schedule page prototype.}

\begin{itemize}
    \item Here is an example of a recommended schedule that the tool would generate based on the information you just provided. 
    \item Do you feel that you understand what the recommended schedule is suggesting?
    \item What part of the recommended schedule do you feel is the most useful? 
    \item What part of the recommended schedule do you feel is the least useful? Is there anything that's missing from the schedule?
    \item Do you feel the recommended schedule gives you enough information to decide whether you would want to follow it?
\end{itemize}

Finally, we'd like to ask about your overall opinion of the tool.
\begin{itemize}
    \item What did you like about this tool?
    \item What did you dislike about this tool?
    \item Did you feel that interacting with the tool took too much time, or that it was too complicated or confusing for you? Why or why not?
    \item What sort of information would increase the chance that you want to use this tool and follow its recommendations? How big of a difference do you think that having this information would make?
    \item Do you believe that drivers would generally find a tool like this to be useful for when they're planning their driving? Why or why not?
\end{itemize}

\subsection{Participants}
\label{sec:app:pilot:participants}
As with the longitudinal user studies (\Cref{sec:userstudy:participants}), participants were recruited from the user base of Gridwise in July 2023. Gridwise distributed recruitment messages to 500 users, but otherwise did not interact with participants. Recipients were sampled from Gridwise users in the United States who had completed at least one gig in a platform linked to the Gridwise app over the week preceding recruitment. We recruited 4 interview participants:
\begin{itemize}
    \item \textbf{P1}: A 39-year-old female with a professional degree who drives exclusively for delivery platforms
    \item \textbf{P2}: A 55-year-old male with less than a high school degree who drives exclusively for delivery platforms
    \item \textbf{P3}: A 29-year-old female with an undergraduate degree who drives more frequently for ridesharing platforms
    \item \textbf{P4}: A 53-year-old male with a professional degree who drives more frequently for delivery platforms
\end{itemize}

\subsection{Results}
\label{sec:app:pilot:results}
The four pilot interview participants reported a diversity of motivations and routines for driving. While all four participants had specific earning goals, P1, P2, and P4 considered their goals to be important and valued the time flexibility of gig work, whereas P3 was more motivated by the opportunity for human interaction. P1, P2 and P3 had typical times that they drive at; however, P1, P2, and P4 also adjusted their schedules based on demand. All four participants had encountered difficulties in planning due to the unpredictability of demand and/or supply (with P1, P2, and P3 feeling that gig platforms provide insufficient information), and indicated that they would find schedule recommendations to be useful.

All four participants found the initial design of the tool to be generally understandable, and felt that it would be useful for drivers in planning their activity. P1 and P2 liked the fact that the tool presents information to them in a way that reduces the need for guesswork while driving. In particular, P1 suggested that the tool would help mitigate a catch-22: it is not possible to view gig demand information in DoorDash without exiting their Dash (scheduled work period), but doing so seemingly deprioritises them.

On the constraint page, P1 and P4 indicated that the questions aligned well with their goals. P1 and P3 suggested that they would not set the constraints to perfectly align with their routines, so as to receive more information from the tool. On the schedule page, all four participants liked the estimated hourly earnings, with P1, P2, and P3 indicating that they would be helpful in deciding whether or not to work at particular times of day. Yet, P2, P3, and P4 acknowledged that the estimates would only be guesses. P1 and P4 also liked the range of weekly earnings, but P3 felt it assumed they would follow the recommended schedule perfectly. P2 and P3 noted that ranges for hourly earnings would be useful to display.

P1, P2, and P3 all felt that it was better for the tool to have a simple, easy-to-use design. All three indicated that the prototype fulfilled this requirement, although P3 suggested that wording and design improvements would be necessary (in particular, they felt that the monotone colour scheme of the tool was confusing). P2 and P3 felt that the design needed to be mobile-friendly. The participants also mentioned other desiderata:
\begin{itemize}
    \item \textbf{More granular constraints.} P1, P3, and P4 all suggested ways to limit the scope of the historical gigs used to estimate their earnings. P3 wanted the tool to clarify that the historical data was limited to the city they drive in, and also how recent the data was. P1 and P3 wanted to limit the maximum distance of the historical gigs from their starting point. P4, who works for less popular delivery platforms, wanted to select which platforms the historical gigs came from, and indicated that this would improve their perceived control.

    \item \textbf{Feedback on performance.} P1 and P2 both wanted a way to compare the tool's estimates with their actual earnings. Regardless of how the estimates compared to reality, P1 suggested that this feedback would be motivating; both P1 and P2 remarked that they have gamified their gig-driving experiences to compare against either themselves or others. P2 also wanted to compare the estimates with their expenses, and P3 wanted a way to view the overall supply of drivers.
\end{itemize}

\section{User Study Interview Script}
\label{sec:app:interview}
\textit{The script below focuses on the guiding questions that we asked participants. Some typical probing questions are also listed as sub-bullets.}

\subsection{Formative Questions}
\label{sec:app:interview:formative}
Let's start with talking about your driving for rideshare/delivery services in general.

\begin{itemize}
    \item Could you start by telling us why you are driving? 
    \begin{itemize}
        \item Is it primary or supplemental income? 
        \item What other commitments do you balance it with (jobs, family, hobbies)?
    \end{itemize}
    \item To what extent do you rely on making a target amount when you are driving? 
    \item Can you talk through your typical process for deciding when to drive? 
\end{itemize}

\subsection{Feedback on Constraint Design}
\label{sec:app:interview:constraint}
Now, let's think back to the times when you were interacting with the tool, particularly when it asked you to enter your availability and goals.

\begin{itemize}
    \item How similar or different were the tool's questions to the way you typically make these decisions?
    \item Did you feel like you were able to use the tool to adequately specify your main considerations for when you'd like to drive? 
    \begin{itemize}
        \item Were you ever unsure of what information the tool was asking for? 
        \item Would you have preferred the tool to ask for information differently, or to ask for different information?
    \end{itemize}
    \item Did you feel like you were able to influence the recommended schedule that the tool generated for you? 
    \begin{itemize}
        \item Did you try to experiment with entering in different information? 
    \end{itemize}
    \item Did you feel that interacting with the tool took too much time, or that it was too complicated or confusing for you? 
    \begin{itemize}
        \item Could you see yourself spending more time interacting with the tool than you did (e.g. to enter more details)? Why or why not?
    \end{itemize}
\end{itemize}

\subsection{Feedback on Schedules}
\label{sec:app:interview:schedules}

Now, let's talk about your how the tool's recommended schedules may or may not have influenced your driving activity over the last few days.

\begin{itemize}
    \item Did you find that the recommended schedules made sense? 
    \begin{itemize}
        \item Did you feel that you understood how the tool used your answers to generate schedules? Why or why not?
    \end{itemize}
    \item To what extent did you rely on the email reminders of the schedules? 
    \begin{itemize}
        \item Did you ever miss the email reminders? 
        \item When did you typically check the schedule, if at all?
    \end{itemize}
    \item If you saw the recommended schedules, how did they impact your process for deciding when to work?
    \begin{itemize}
        \item To what extent did you follow the schedules? 
        \item Were there times at which you prioritised your own intuition over the schedules? 
        \item If so, were there times at which you wished you followed the schedule more closely? Why or why not?
    \end{itemize} 
    \item How did your response to recommended schedules change throughout the week, if at all? 
    \begin{itemize}
        \item Did you look at the schedules more or less as time went on? 
        \item Were there any particular days on which you wanted to check the schedule more? Why or why not?
    \end{itemize}
    \item Did you feel that the tool gave you more or different information than you would otherwise get from the services that you drive for/from Gridwise? Why or why not?
    \item Are there some additional details which could have increased the chance that you followed the recommended schedules? 
    \begin{itemize}
        \item For example, would you have preferred to see the schedule for the entire week on every day?
    \end{itemize}
    \item Did the recommended schedules lead you to drive at different times and/or locations than before? 
    \begin{itemize}
        \item Did this happen early on or later? 
        \item At what times of day?
    \end{itemize}
    \item When you followed the recommended schedules, did you feel that you ended up making more money, less money, or about the same relative to before? 
    \begin{itemize}
        \item How closely do you track your earnings in general? 
        \item Did you track your earnings more closely when using the tool?
    \end{itemize}
\end{itemize}

\subsection{Feedback on Estimates}
\label{sec:app:interview:estimates}

\begin{itemize}
    \item How much did you focus on the tool's estimates for how much you could earn?
    \item Did you feel like you could rely on the estimates to achieve your earning goals? 
    \begin{itemize}
        \item Did you feel that these estimates were meant to be accurate projections of how much you could earn, or that they were rough ballpark figures?
    \end{itemize}
    \item In general, did you feel that the estimated earnings had the right level of detail, or would you have liked to see additional information?
    \begin{itemize}
        \item \textit{[If participants were in Conditions~(B) or (D)]} Would you have preferred to see a range for how much you could earn?
        \item \textit{[If participants were in Conditions~(R) or (RH)]} Would you have preferred to see a single number for how much you could earn?
    \end{itemize}
\end{itemize}

\textit{The interviewer selects a particular day on which the participant interacted with the tool. If earnings data was available, this was a day on which the participant earned more than the tool's estimate; otherwise, this was the sixth day of their interaction with the tool.}

\begin{itemize}
    \item Let's talk about [weekday], Day [day] of your interaction with the tool. 
    \item Do you recall the extent to which you looked at the recommended schedule? 
    \begin{itemize}
        \item If you did, do you remember how you decided whether you wanted to follow it? Was this influenced by how much the tool estimated your earnings to be? Why or why not? What did you think of the estimate that the tool gave you?
        \item If you didn't, do you happen to remember why? Was this influenced by how much you earned on the previous day? Why or why not?
    \end{itemize} 
    \item Do you recall whether you made more or less than the tool estimated on that day? 
    \begin{itemize}
        \item If there were any differences, do you have any idea why?
    \end{itemize} 
    \item Did that influence your decision to look at the recommended schedule for the next day? Why or why not?
    \item We checked your records briefly and found that you earned \$xxx.xx, compared to the tool's estimate of \$xxx.xx. Does that change how you feel at all?
\end{itemize}

\subsection{Overall Thoughts}
\label{sec:app:interview:overall}
Now, we'd like to wrap up with a few general questions about the tool.
\begin{itemize}
    \item Did you feel that the time you spent interacting with the tool was worthwhile or not worthwhile? Why or why not?
    \item If you had the option of using a tool like this one, what are the chances that you might actually use it to decide your driving schedule in the future? Why or why not?
    \item Beyond what you've mentioned already, is there anything else you believe might increase the chance that you would use this tool in the future?
    \item Do you have any other questions, comments, or concerns?
\end{itemize}

\section{Full Quantitative Results}
\label{sec:app:quant}

\subsection{Trust and Reliance Score LMMs}
\label{sec:app:quant:lmms}
In \Cref{tab:trust-model} and \Cref{tab:reliance-model}, we show the fitted coefficients for the trust and reliance score LMMs discussed in \Cref{sec:quant:longitudinal} and \Cref{sec:quant:condition:longitudinal}.

\subsection{Daily Trust and Reliance Statistics}
\label{sec:app:quant:daily-stats}
\begin{figure*}[ht]
    \centering
    \includegraphics[width=0.8\linewidth]{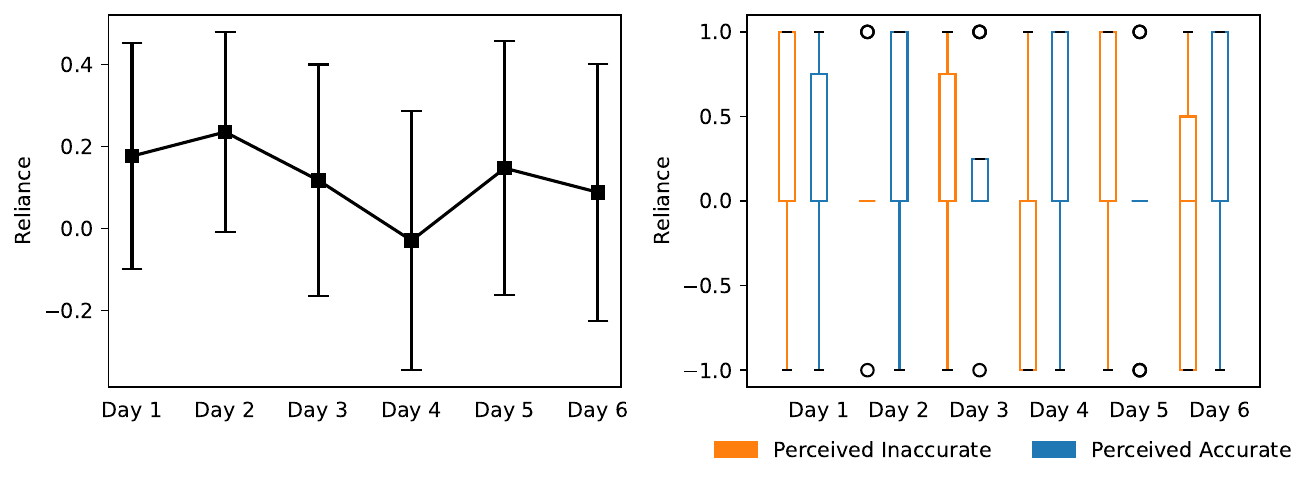}
    \caption{(L) Means and 95\% CIs of reliance scores for the schedule recommendation tool on Days 1--6 among retained participants. Full statistics are shown in \Cref{tab:daily-stats}. (R) Boxplots of reliance scores on Days 1--6, decomposed by perceived accuracy.}
    \label{fig:daily-reliance}
    \Description{On the left, a line plot with seven points decreases from 0.176 on Day 1 to 0.088 on Day 6. The general trend is downwards but not monotonic. 95\% confidence intervals for each of the points are overlapping. On the right, a boxplot shows two boxes for each day: one red (perceived inaccurate) and one green (perceived accurate). There is no clear trend in the IQRs, which all overlap. Most of the IQRs are between 0 and 1, but some of them consist of 0 only.}
\end{figure*}

In \Cref{fig:daily-reliance}, we show visualisations for daily reliance scores omitted from the main text. In \Cref{tab:daily-stats}, we report the means, standard errors, and 95\% confidence intervals of the daily trust and reliance scores plotted in \Cref{fig:daily-trust,fig:daily-reliance}. We show both the overall statistics as well as the statistics for each condition.

\begin{table*}[ht]
    \centering
    \begin{tabular}{lSSlSSl}
        \toprule
        \multirow{2}{*}{\textbf{Factor}} & \multicolumn{3}{c}{\textbf{Without Condition}} & \multicolumn{3}{c}{\textbf{With Condition}} \\
        \cmidrule(lr){2-4} \cmidrule(lr){5-7}
        & \multicolumn{1}{c}{$\beta$} & \multicolumn{1}{c}{\textit{SE}} & \multicolumn{1}{c}{$p$} & \multicolumn{1}{c}{$\beta$} & \multicolumn{1}{c}{\textit{SE}} & \multicolumn{1}{c}{$p$} \\
        \midrule
        Intercept                       &  1.447 & 0.468 &  0.00337**  &   1.994 & 0.549 & 0.00068*** \\
        \texttt{pre\_trust\_score}      &  0.471 & 0.119 &  0.00029*** &   0.411 & 0.128 & 0.00256**  \\
        \texttt{day}                    &  0.130 & 0.027 & <0.00001*** &   0.052 & 0.046 & 0.25896    \\
        \texttt{estimate\_accurate}     &  0.415 & 0.167 &  0.01357*   &   0.392 & 0.167 & 0.02010*   \\
        \texttt{day:estimate\_accurate} & -0.121 & 0.037 &  0.00126**  &  -0.119 & 0.037 & 0.00168**  \\
        \texttt{condition(D)}           &        &       &             &  -0.494 & 0.341 & 0.15139    \\
        \texttt{condition(R)}           &        &       &             &  -0.169 & 0.360 & 0.64038    \\
        \texttt{condition(RH)}          &        &       &             &  -0.487 & 0.341 & 0.15755    \\
        \texttt{day:condition(D)}       &        &       &             &   0.111 & 0.053 & 0.03548*   \\
        \texttt{day:condition(R)}       &        &       &             &   0.103 & 0.056 & 0.06976\textdagger \\
        \texttt{day:condition(RH)}      &        &       &             &   0.075 & 0.052 & 0.14971    \\
        Random intercept SD             &  0.605 &       &             &   0.613 &       &            \\
        \bottomrule
    \end{tabular}
    \caption{Factors and coefficients ($\beta$ with standard error $SE$) for our linear mixed model of daily trust scores, without and with the \texttt{condition} as an independent variable. Statistically significant coefficients are denoted as \textdagger~(0.1), *~(0.05), **~(0.01), ***~(0.001).}
    \label{tab:trust-model}
\vspace{-2em}
\end{table*}

\begin{table*}[ht]
    \centering
    \begin{tabular}{lSSlSSl}
        \toprule
        \multirow{2}{*}{\textbf{Factor}} & \multicolumn{3}{c}{\textbf{Without Condition}} & \multicolumn{3}{c}{\textbf{With Condition}} \\
        \cmidrule(lr){2-4} \cmidrule(lr){5-7}
        & \multicolumn{1}{c}{$\beta$} & \multicolumn{1}{c}{\textit{SE}} & \multicolumn{1}{c}{$p$} & \multicolumn{1}{c}{$\beta$} & \multicolumn{1}{c}{\textit{SE}} & \multicolumn{1}{c}{$p$} \\
        \midrule
        Intercept                       & -0.024 & 0.422 & 0.95494  & -0.129 & 0.496  & 0.79607  \\
        \texttt{pre\_trust\_score}      &  0.104 & 0.086 & 0.23316  &  0.103 & 0.092  & 0.27397  \\
        \texttt{day}                    & -0.220 & 0.089 & 0.01444* & -0.234 & 0.102  & 0.02280* \\
        \texttt{estimate\_accurate}     &  0.124 & 0.186 & 0.50517  &  0.125 & 0.187  & 0.50469  \\
        \texttt{trust\_score}           & -0.115 & 0.099 & 0.24285  & -0.088 & 0.101  & 0.38470  \\
        \texttt{day:estimate\_accurate} &  0.012 & 0.048 & 0.81042  &  0.008 & 0.049  & 0.86786  \\
        \texttt{day:trust\_score}       &  0.058 & 0.024 & 0.01640* &  0.045 & 0.025  & 0.07127\textdagger \\
        \texttt{condition(D)}           &        &       &          &  0.028 & 0.294  & 0.92461  \\
        \texttt{condition(R)}           &        &       &          & -0.145 & 0.308  & 0.63782  \\
        \texttt{condition(RH)}          &        &       &          &  0.139 & 0.291  & 0.63414  \\
        \texttt{day:condition(D)}       &        &       &          &  0.062 & 0.067  & 0.35761  \\
        \texttt{day:condition(R)}       &        &       &          &  0.137 & 0.072  & 0.05958\textdagger \\
        \texttt{day:condition(RH)}      &        &       &          &  0.043 & 0.065  & 0.50832  \\
        Random intercept SD             &  0.366 &       &          &  0.382 &        &          \\
        \bottomrule
    \end{tabular}
    \caption{Factors and coefficients ($\beta$ with standard error $SE$) for our linear mixed model of daily reliance scores, without and with the \texttt{condition} as an independent variable. Statistically significant coefficients are denoted as \textdagger~(0.1), *~(0.05), **~(0.01), ***~(0.001).}
    \label{tab:reliance-model}
\vspace{-2em}
\end{table*}

\newpage
\onecolumn
\subsection{Post-Interaction Trust}
\label{sec:app:quant:post}
We fitted an ordinary least squares (OLS) model for the \texttt{post\_trust\_score} using \texttt{statsmodels}. This model included the \texttt{condition} along with all previous trust (\texttt{pre\_trust\_score} and daily \texttt{trust\_score}) and \texttt{reliance} measurements.
\begin{align*}
    \texttt{post\_trust\_score} \sim &\ \texttt{condition} + \texttt{pre\_trust\_score} + \sum_{i=1}^7 (\texttt{trust\_score\_$i$} + \texttt{reliance\_$i$})
\end{align*}
For Conditions~(B)/(D)/(R)/(RH), the mean post-interaction trust scores were 4.000, 3.720, 3.857, and 3.618. None of these conditions were significantly different from each other, and the coefficients for previous trust and reliance scores were not statistically significant either. The questions we adapted from the TiA asked participants to consider the entire duration of their interaction with the schedule recommendation tool. This broad, retrospective reflection may have failed to capture more nuanced longitudinal changes in trust and reliance like those we described in \Cref{sec:quant}.

\begin{table*}[ht]
    \centering
    \begin{tabular}{cccccccc}
        \toprule
        \multirow{2}{*}{\textbf{Day}} & \multirow{2}{*}{\textbf{Condition}} & 
        \multicolumn{3}{c}{\textbf{Trust}} & \multicolumn{3}{c}{\textbf{Reliance}} \\
        \cmidrule(lr){3-5} \cmidrule(lr){6-8}
        & & $\mu$ & \textit{SE} & 95\% CI & $\mu$ & \textit{SE} & 95\% CI \\
        \midrule
        \multirow{5}{*}{1} & Overall & 3.353 & 0.196 & (2.872, 3.834) &  0.176 & 0.107 & (-0.100, 0.453) \\
                           & (B)     & 3.556 & 0.444 & (2.468, 4.643) &  0.167 & 0.307 & (-0.623, 0.957) \\
                           & (D)     & 3.033 & 0.390 & (2.080, 3.986) &  0.200 & 0.200 & (-0.314, 0.714) \\
                           & (R)     & 3.810 & 0.348 & (2.959, 4.660) &  0.143 & 0.261 & (-0.528, 0.813) \\
                           & (RH)    & 3.242 & 0.379 & (2.315, 4.170) &  0.182 & 0.182 & (-0.286, 0.649) \\
        \midrule
        \multirow{5}{*}{2} & Overall & 3.578 & 0.167 & (3.169, 3.988) &  0.235 & 0.095 & (-0.009, 0.479) \\
                           & (B)     & 3.722 & 0.416 & (2.703, 4.741) &  0.333 & 0.333 & (-0.524, 1.190) \\
                           & (D)     & 3.567 & 0.205 & (3.065, 4.069) &  0.200 & 0.200 & (-0.314, 0.714) \\
                           & (R)     & 3.857 & 0.397 & (2.885, 4.830) &  0.429 & 0.202 & (-0.091, 0.948) \\
                           & (RH)    & 3.333 & 0.362 & (2.447, 4.220) &  0.091 & 0.091 & (-0.143, 0.325) \\
        \midrule
        \multirow{5}{*}{3} & Overall & 3.500 & 0.168 & (3.089, 3.911) &  0.118 & 0.110 & (-0.165, 0.400) \\
                           & (B)     & 3.833 & 0.331 & (3.025, 4.642) & -0.167 & 0.307 & (-0.957, 0.623) \\
                           & (D)     & 3.400 & 0.364 & (2.508, 4.292) &  0.100 & 0.180 & (-0.361, 0.561) \\
                           & (R)     & 4.000 & 0.309 & (3.245, 4.755) &  0.571 & 0.202 & ( 0.052, 1.091) \\
                           & (RH)    & 3.091 & 0.270 & (2.430, 3.752) &  0.000 & 0.191 & (-0.490, 0.490) \\
        \midrule
        \multirow{5}{*}{4} & Overall & 3.755 & 0.148 & (3.392, 4.118) & -0.029 & 0.123 & (-0.346, 0.287) \\
                           & (B)     & 3.778 & 0.306 & (3.028, 4.527) & -0.167 & 0.307 & (-0.957, 0.623) \\
                           & (D)     & 3.667 & 0.157 & (3.282, 4.051) & -0.200 & 0.200 & (-0.714, 0.314) \\
                           & (R)     & 4.238 & 0.347 & (3.390, 5.086) &  0.286 & 0.286 & (-0.449, 1.020) \\
                           & (RH)    & 3.515 & 0.334 & (2.697, 4.333) &  0.000 & 0.234 & (-0.600, 0.600) \\
        \midrule
        \multirow{5}{*}{5} & Overall & 3.588 & 0.159 & (3.199, 3.978) &  0.147 & 0.120 & (-0.162, 0.457) \\  
                           & (B)     & 3.278 & 0.475 & (2.116, 4.439) & -0.167 & 0.307 & (-0.957, 0.623) \\  
                           & (D)     & 3.733 & 0.257 & (3.104, 4.363) &  0.100 & 0.180 & (-0.361, 0.561) \\  
                           & (R)     & 4.143 & 0.290 & (3.434, 4.852) &  0.429 & 0.297 & (-0.336, 1.193) \\  
                           & (RH)    & 3.273 & 0.273 & (2.605, 3.940) &  0.182 & 0.226 & (-0.400, 0.764) \\  
        \midrule
        \multirow{5}{*}{6} & Overall & 3.667 & 0.163 & (3.268, 4.065) &  0.088 & 0.122 & (-0.226, 0.402) \\
                           & (B)     & 3.333 & 0.487 & (2.142, 4.525) & -0.333 & 0.333 & (-1.190, 0.524) \\
                           & (D)     & 3.733 & 0.276 & (3.059, 4.408) &  0.200 & 0.200 & (-0.314, 0.714) \\
                           & (R)     & 4.143 & 0.340 & (3.311, 4.975) &  0.714 & 0.184 & ( 0.240, 1.188) \\
                           & (RH)    & 3.485 & 0.275 & (2.813, 4.157) & -0.182 & 0.182 & (-0.649, 0.286) \\
        \midrule
        \multirow{5}{*}{7} & Overall & 3.843 & 0.149 & (3.478, 4.208) & N/A & N/A & N/A \\
                           & (B)     & 3.889 & 0.351 & (3.029, 4.749) & N/A & N/A & N/A \\
                           & (D)     & 3.733 & 0.247 & (3.128, 4.339) & N/A & N/A & N/A \\
                           & (R)     & 4.381 & 0.286 & (3.682, 5.080) & N/A & N/A & N/A \\
                           & (RH)    & 3.576 & 0.292 & (2.862, 4.289) & N/A & N/A & N/A \\
        \bottomrule
    \end{tabular}
    \caption{Statistics for daily trust and reliance, as measured by End-of-Day Surveys.}
    \label{tab:daily-stats}
\end{table*}
\clearpage

\newpage
\section{Figma Prototype Design}
\label{sec:app:figma}
The following figures show screenshots of the Figma prototype used for the pilot interviews (\Cref{sec:app:pilot}).

\clearpage

\newpage
\section{Final Tool Design}
\label{sec:app:tool}

The following figures show screenshots of the web-based schedule recommendation tool used for the longitudinal user studies (\Cref{sec:userstudy}), showing the constraints and recommended schedule of interview participant P1.

\begin{figure*}[ht]
    \centering
    \includegraphics[width=0.825\linewidth]{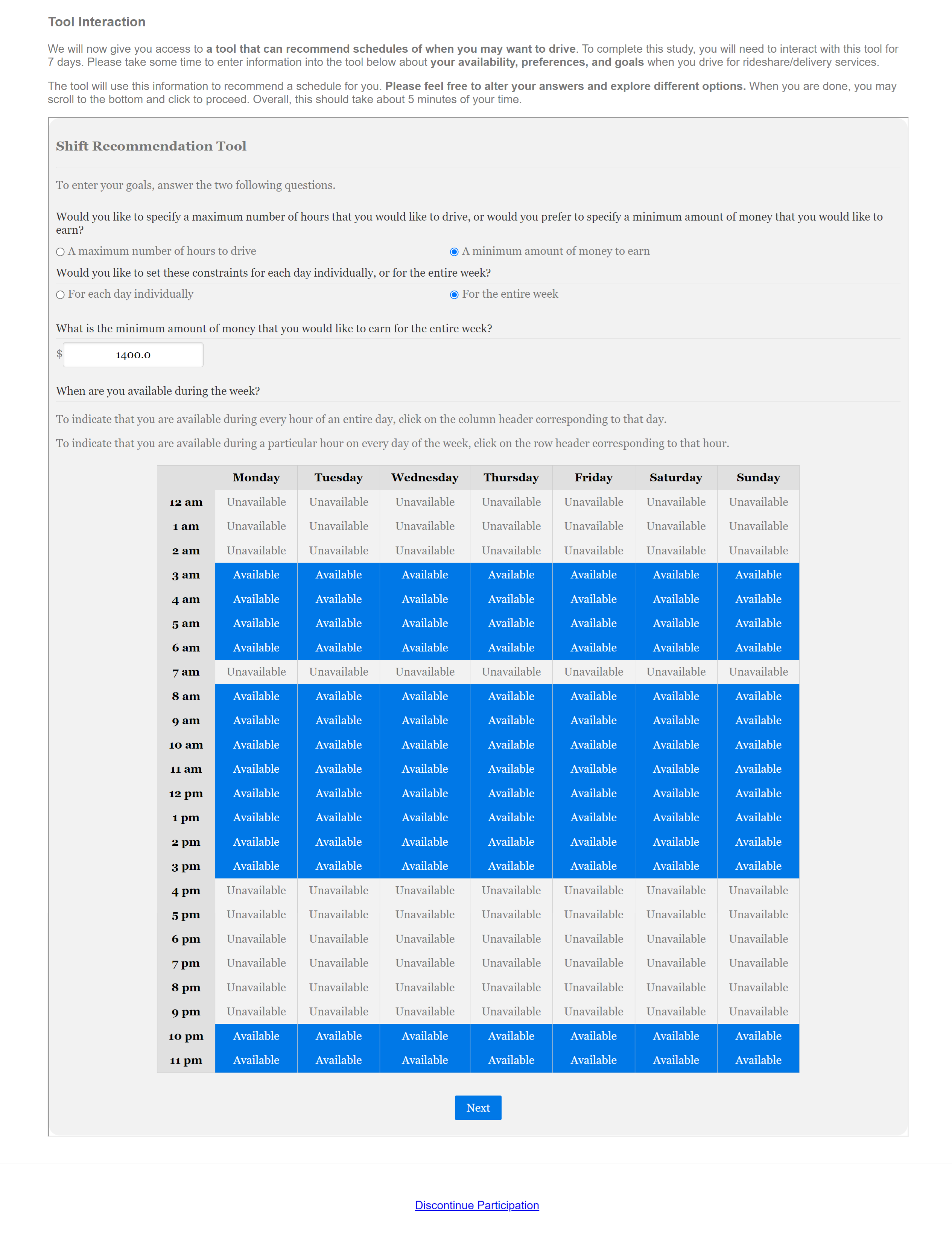}
    \caption{Screenshot of final constraint page, showing constraints entered by interview participant P1.}
    \label{fig:tool:constraint}
    \Description{Screenshot of a webpage titled "Tool Interaction". Below a set of instructions, the constraint page is shown in a grey iframe. Various inputs for constraints are set to values entered by interview participant P1. In response to "Would you like to specify a maximum number of hours that you would like to drive, or would you prefer to specify a minimum amount of money that you would like to earn?", P1 selected the latter. In response to "Would you like to set these constraints for each day individually, or for the entire week?", P1 selected the latter. In response to "What is the minimum amount of money you would like to earn for the entire week?", P1 entered \$1400. In response to "When are you available during the week?", P1 selected time slots (as indicated by blue cells in a tabular schedule) corresponding to 3 am to 7 am, 8 am to 4 pm, and 10 pm to 12 am on every day of the week.}
\end{figure*}

\subsection{Schedule Page Conditions}
\label{sec:app:tool:conditions}
\begin{figure*}[ht]
    \centering
    \includegraphics[width=0.725\linewidth]{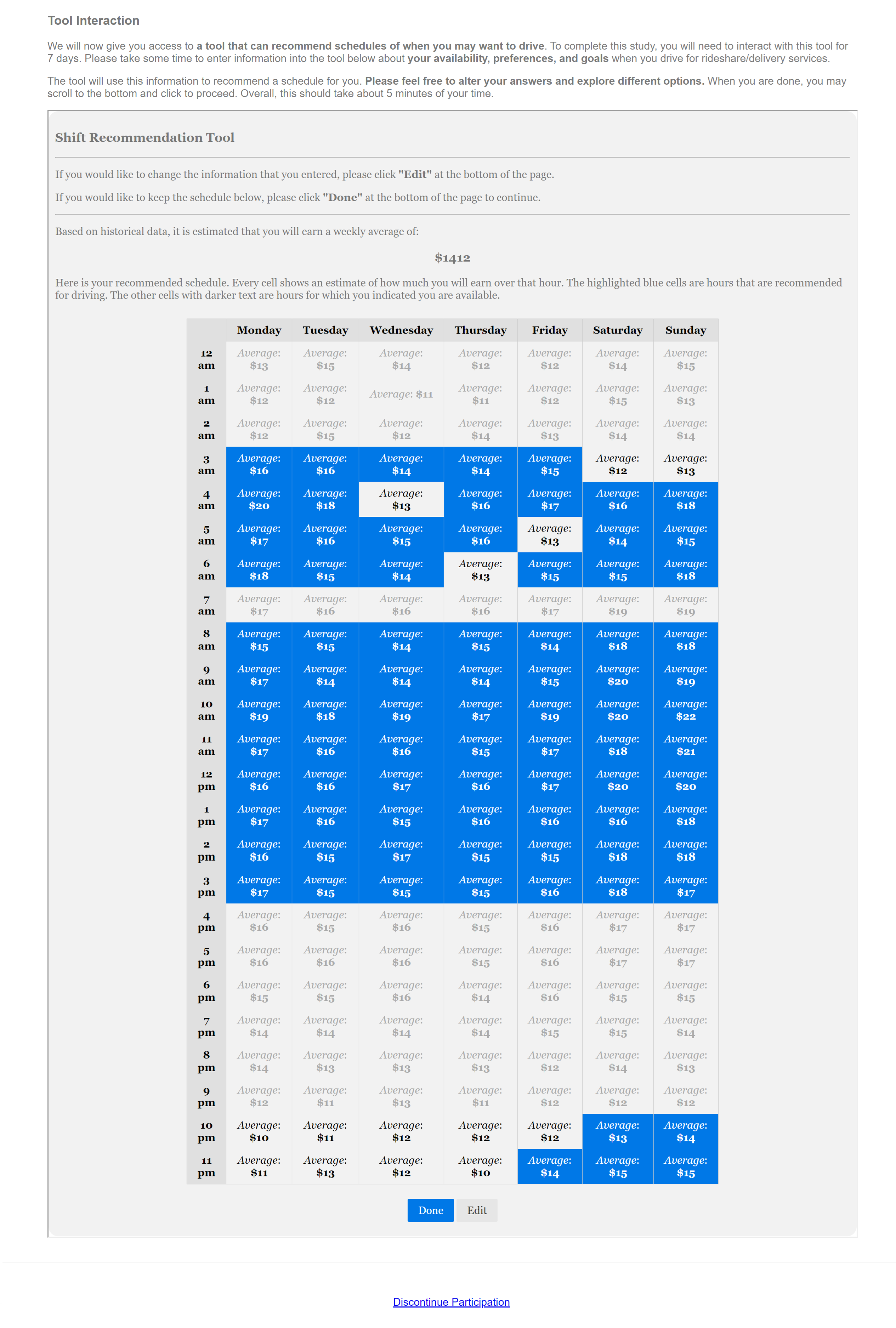}
    \caption{Screenshot of final schedule page, showing the recommended schedule for interview participant P1 as it would be displayed if they were placed in Condition~(B) (base condition; see \Cref{sec:tool:conditions}).}
    \label{fig:tool:schedule-group-1}
    \Description{Screenshot of a webpage titled "Tool Interaction". Below a set of instructions, the schedule page in Condition (B) for interview participant P1 is shown in a grey iframe. The top of the schedule page reads, "Based on historical data, it is estimated that you will earn a weekly average of", followed by an estimate of \$1412 below. Further below, a tabular schedule is shown for each day of the week. Each cell reads, "Average", followed by the estimated hourly earning in bold. Recommended time slots are indicated by blue cells, non-recommended time slots during which P1 is still available are indicated by dark text, and time slots during which P1 is not available are indicated by light text.}
\end{figure*}

\clearpage
\begin{figure*}[ht]
    \centering
    \includegraphics[width=0.725\linewidth]{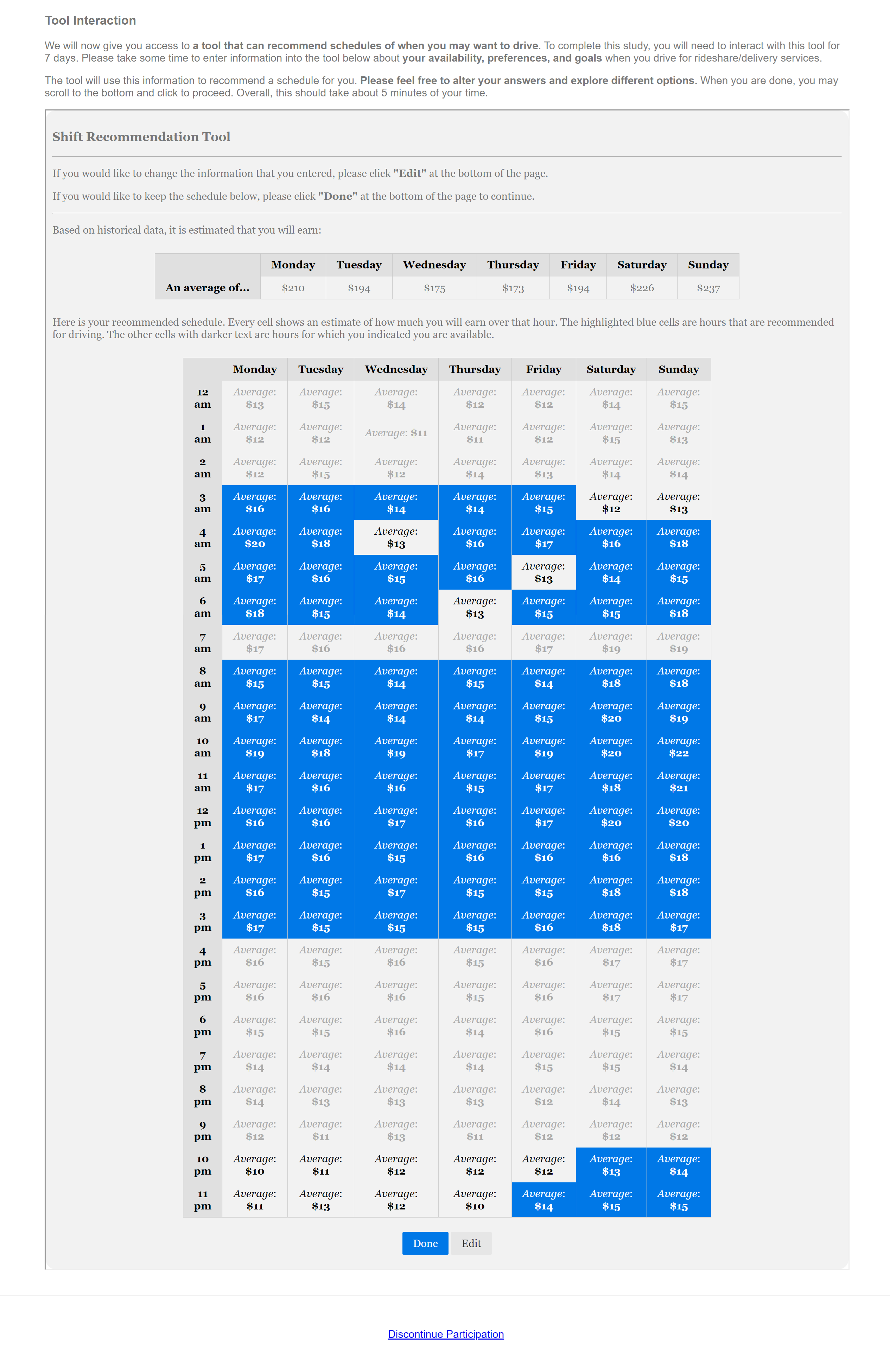}
    \caption{Screenshot of final schedule page, showing the recommended schedule for interview participant P1 as it would be displayed if they were placed in Condition~(D) (daily estimates; see \Cref{sec:tool:conditions}).}
    \label{fig:tool:schedule-group-2}
    \Description{Screenshot of a webpage titled "Tool Interaction". Below a set of instructions, the schedule page in Condition (D) for interview participant P1 is shown in a grey iframe. The top of the schedule page reads, "Based on historical data, it is estimated that you will earn". Unlike Condition (B), this is followed by a table that reads, "An average of", alongside estimates for each day: \$210 for Monday, \$194 for Tuesday, \$175 for Wednesday, \$173 for Thursday, \$194 for Friday, \$226 for Saturday, and \$237 for Sunday. Further below, a tabular schedule is shown for each day of the week. Each cell reads, "Average", followed by the estimated hourly earning in bold. Recommended time slots are indicated by blue cells, non-recommended time slots during which P1 is still available are indicated by dark text, and time slots during which P1 is not available are indicated by light text.}
\end{figure*}

\clearpage
\begin{figure*}[ht]
    \centering
    \includegraphics[width=0.5\linewidth]{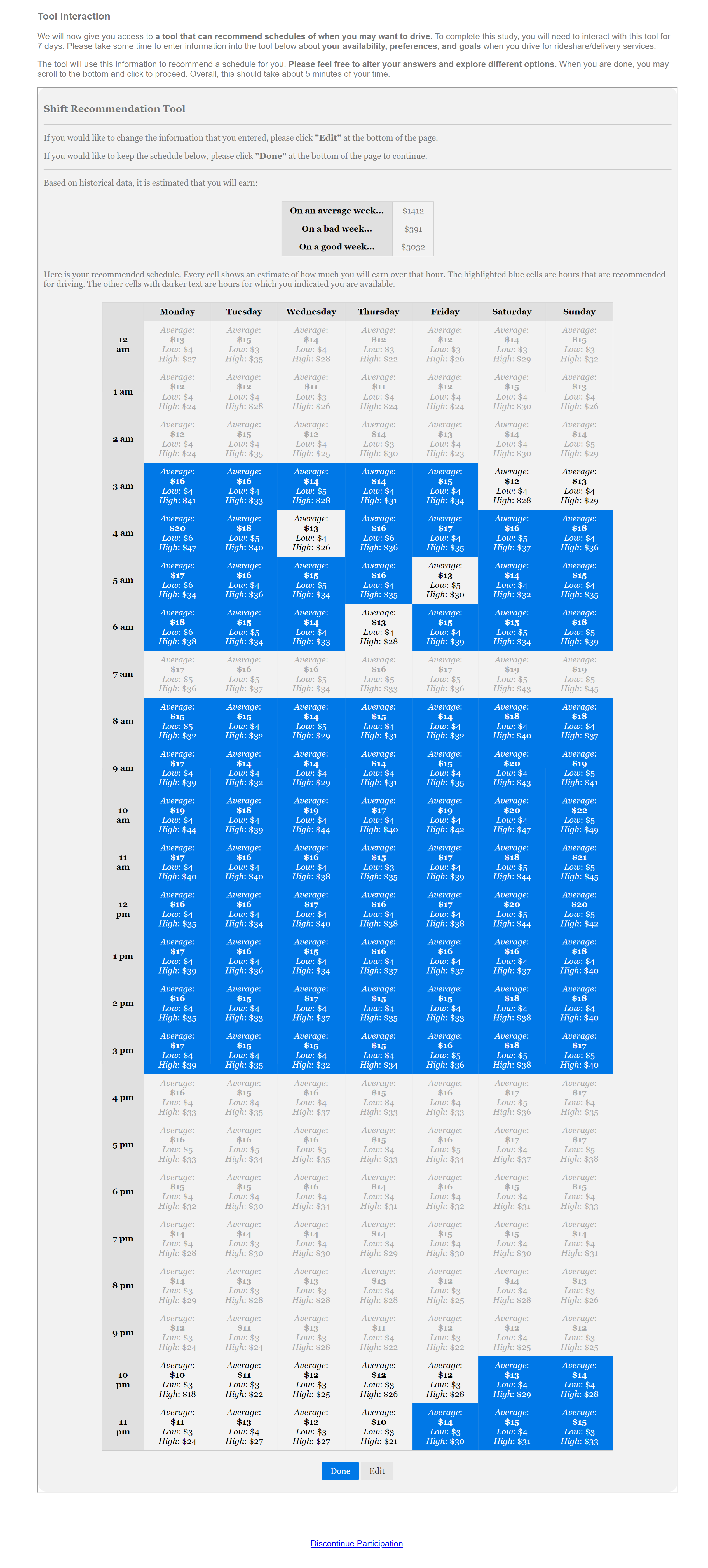}
    \caption{Screenshot of final schedule page, showing the recommended schedule for interview participant P1 as it would be displayed if they were placed in Condition~(R) (ranged estimates; see \Cref{sec:tool:conditions}).}
    \label{fig:tool:schedule-group-3}
    \Description{Screenshot of a webpage titled "Tool Interaction". Below a set of instructions, the schedule page in Condition (R) for interview participant P1 is shown in a grey iframe. The top of the schedule page reads, "Based on historical data, it is estimated that you will earn". Unlike Conditions (B) and (D), this is followed by a table that reads, "On an average week: \$1412", "On a bad week: \$391", and "On a good week: \$3032". Further below, a tabular schedule is shown for each day of the week. Each cell reads, "Average", followed by the estimated hourly earning in bold. Unlike Conditions (B) and (D), this is also followed by "Low", with the 10th percentile of earnings, and "High", with the 90th percentile of earnings. Recommended time slots are indicated by blue cells, non-recommended time slots during which P1 is still available are indicated by dark text, and time slots during which P1 is not available are indicated by light text.}
\end{figure*}

\clearpage
\begin{figure*}[ht]
    \centering
    \includegraphics[width=0.5\linewidth]{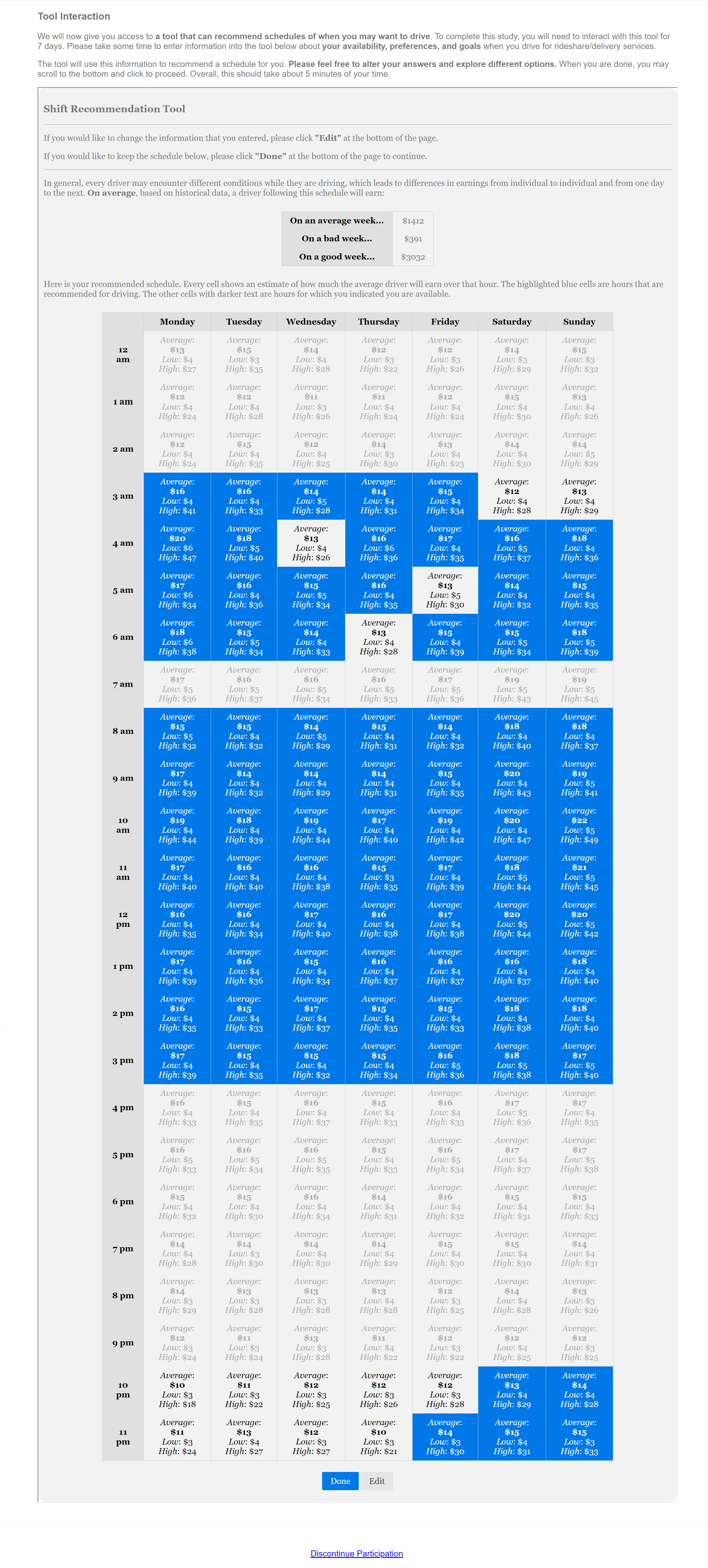}
    \caption{Screenshot of final schedule page, showing the recommended schedule for interview participant P1 as it would be displayed if they were placed in Condition~(RH) (ranged and hedged estimates; see \Cref{sec:tool:conditions}).}
    \label{fig:tool:schedule-group-4}
    \Description{Screenshot of a webpage titled "Tool Interaction". Below a set of instructions, the schedule page in Condition (RH) for interview participant P1 is shown in a grey iframe. Unlike Conditions (B), (D), and (R), the top of the schedule page reads, "In general, every driver may encounter different conditions while they are driving, which leads to differences in earnings from individual to individual and from one day to the next. On average, based on historical data, a driver following this schedule will earn". Also unlike Conditions~(B) and (D), this is followed by a table that reads, "On an average week: \$1412", "On a bad week: \$391", and "On a good week: \$3032". Further below, a tabular schedule is shown for each day of the week. Each cell reads, "Average", followed by the estimated hourly earning in bold. Unlike Conditions (B) and (D), this is also followed by "Low", with the 10th percentile of earnings, and "High", with the 90th percentile of earnings. Recommended time slots are indicated by blue cells, non-recommended time slots during which P1 is still available are indicated by dark text, and time slots during which P1 is not available are indicated by light text.}
\end{figure*}

\end{document}